\documentclass[fleqn,usenatbib]{rasti}

\usepackage{newtxtext,newtxmath}

\usepackage[T1]{fontenc}

\DeclareRobustCommand{\VAN}[3]{#2}
\let\VANthebibliography\thebibliography
\def\thebibliography{\DeclareRobustCommand{\VAN}[3]{##3}\VANthebibliography}

\usepackage{graphicx}	
\usepackage{amsmath}	
\usepackage{subfig}
\usepackage{comment} 

\title[Characterizing Stellar Flares in Ariel Targets]{Characterizing Stellar Flares in Ariel Targets: Activity Analysis and Transit Contamination}

\author[G. Galletta et al.]{
G. Galletta,$^{1,2}$\thanks{E-mail: gabriele.galletta@inaf.it}
S. Colombo,$^{3}$
and G. Micela$^{3}$
\\
$^{1}$Dipartimento di Fisica e  Chimica, Università di Palermo, Via Archirafi 36, 90128, Palermo, Italy.\\
$^{2}$Blue Skies Space Italia S.R.L. Via Vincenzo Monti 16, 20123, Milano, Italy.\\
$^{3}$INAF, Osservatorio Astronomico di Palermo, Palermo, Italy.
}

\date{Accepted 2026 August 16. Received 2026 July 08; in original form 2026 February 23}

\pubyear{\the\year{}}

\begin{document}
\label{firstpage}
\pagerange{\pageref{firstpage}--\pageref{lastpage}}
\maketitle

\begin{abstract}
Stellar flares are sudden releases of magnetic energy that can distort exoplanet transit photometry and transmission spectroscopy, biasing planet radius estimates, transit timings, and atmospheric characterization. Understanding flare activity in Ariel targets is therefore essential to identify stars where flares may compromise observations and to characterize the radiation environment affecting atmospheric escape and photochemistry. We analyzed 290 Ariel target stars using TESS light curves. Flares were identified via iterative Gaussian process detrending, and their energy distributions were modeled with two-segment power laws. We performed injection–recovery tests by adding synthetic flares to detrended light curves and running the full pipeline to quantify completeness and detection biases. We detected 15,857 flares across 1,638 TESS sectors, with 2–86 events per sector. We defined a normalized flare index GF.01 to compare activity across stellar luminosities. Near 3\% of the sample exhibits enhanced flare activity (GF.01 > 1). AU Mic and HD 28109 show a high likelihood of flare contamination during transit observations. GF.01 correlates negatively with stellar bolometric luminosity, indicating higher relative flare output in lower-luminosity stars. AU Mic is an extreme case: four of five observed transits of AU Mic b are affected by flares, consistent with statistical expectations. We validate the framework by comparing predicted flare-contamination probabilities with observed flare occurrences in a representative subset of transits, finding agreement within uncertainties. These results confirm that energetic flares can significantly impact transit observations and provide quantitative guidance for Ariel target selection and analysis strategies.
\end{abstract}




\section{Introduction}\label{sec:Introduction}
The brightness of solar-type and smaller stars is intrinsically variable on a wide range of temporal and energetic scales, from short, impulsive events lasting seconds to long-term magnetic cycles spanning years–decades \citep[e.g.][]{radick1990stellar}. This variability arises from multiple, physically distinct processes, including surface convection and the emergence of photospheric spots driven by the stellar dynamo \citep[e.g.][]{donati2008magnetic}, quasi-periodic activity cycles \citep[e.g.][]{jeffers2023stellar}, and impulsive releases of magnetic energy through reconnection \citep[e.g.][]{kowalski2024stellar}. In some systems, interactions between the star and close-in planets can further modulate activity, producing detectable signatures that differ from purely stellar-driven behaviour \citep[e.g.][]{cauley2019magnetic}.\\
Among stellar activity phenomena, flares are particularly important: they are sudden, large releases of magnetic energy that produce intense bursts of electromagnetic radiation across the spectrum on short timescales \citep[solar analogues were discussed early on by][]{carmichael196454}. Because flares probe the processes of magnetic-field generation and dissipation, their occurrence rates, energies and temporal profiles provide direct constraints on dynamo operation and magnetic reconnection in stars of different types and ages \citep[e.g.][]{donati2008magnetic,lanza2009stellar}.\\
Stellar variability, and flares in particular, has multiple direct implications for exoplanet science. Surface heterogeneities, including spots, faculae, and flares, can distort both transit photometry and transmission spectroscopy, potentially biasing inferred planet radii, transit timings, and atmospheric signatures if these effects are not properly accounted for \citep[][]{ballerini2012multiwavelength}. This highlights the importance of incorporating stellar activity models and flare statistics when interpreting exoplanet observations, especially for active stars. In particular, both occulted and unocculted stellar heterogeneities can imprint wavelength-dependent signals on transmission spectra, giving rise to the Transit Light Source Effect (TLSE), which can mimic or obscure planetary atmospheric features \citep{rackham2017access,rackham2018transit}. For example, simulations by \cite{oshagh2013effect} show that stellar spots may lead to underestimates of the planet radius by $\sim$4 \%, errors of transit duration of similar magnitude, and even transit-timing variations (TTVs) up to $\sim$200 seconds, all derivable from anomalies in the light-curve caused by occulted or non-occulted spots. Similarly, stellar flares occurring during transits can distort light curves and contaminate transmission spectra, potentially leading to incorrect estimates of transit parameters or misinterpretation of atmospheric features if not properly modeled \citep[e.g.][]{howard2023characterizing}.\\ Beyond their observational impact, flares can also physically affect exoplanetary atmospheres. High-energy radiation and particle fluxes associated with flares can change atmospheric chemistry (e.g. ozone and other key species), drive enhanced photochemistry, and contribute to atmospheric escape, thereby modifying the long-term habitability and observability of planetary atmospheres \citep[e.g.][]{buccino2007uv,segura2010effect,johnstone2016influences}. For these reasons, quantifying flare activity is essential both to interpret exoplanet observations reliably and to assess the environmental impact of stellar activity on planets.\\
A substantial body of work has characterized stellar flare statistics using space-based photometric surveys. Early large-scale flare studies with the Kepler mission revealed that flare frequency distributions (FFDs) follow power-law behaviours spanning several orders of magnitude in energy, and that flare rates strongly depend on stellar rotation and spectral type \citep{shibayama2013superflares,davenport2014kepler}. These studies demonstrated that both Sun-like and low-mass stars can produce rare but extremely energetic superflares, establishing a statistical framework for flare occurrence.
More recently, the TESS mission has significantly expanded flare studies to a much larger and more diverse stellar sample. Systematic analyses have provided flare catalogs for hundreds to thousands of stars, confirming the general power-law behaviour of FFDs while extending them to different stellar populations and ages \citep{gunther2020stellar,howard2019evryflare,howard2022flaring,ilin2021flares}. In particular, M-dwarf stars have been shown to exhibit the highest flare rates and energies, making them key targets for studies of stellar magnetic activity and its impact on exoplanet environments \citep{davenport2014kepler,gunther2020stellar,ilin2021flares}.
These observational results provide the empirical basis for modeling flare occurrence rates and for interpreting flare contamination in exoplanet transit and atmospheric studies. These studies motivate the need for homogeneous flare analyses across stellar samples, such as the one presented in this work.\\
The Atmospheric Remote-sensing Infrared Exoplanet Large-survey (Ariel) is the fourth medium-class mission (M4) in ESA’s Cosmic Vision programme, scheduled for launch in 2031 \citep[][]{tinetti2016science, tinetti2018chemical, 2022EPSC...16.1114T}. It is the first space observatory specifically designed to perform a large and homogeneous survey of exoplanet atmospheres, aiming to observe about one thousand transiting planets across a wide range of stellar types and orbital configurations. By combining transit, eclipse, and phase-curve spectroscopy, Ariel will provide precise measurements of atmospheric composition, thermal structure, and cloud properties, thereby enabling a comparative understanding of planetary formation and evolution beyond the Solar System.\\
Understanding the flaring activity of stellar targets is critical for the success of the Ariel mission since stellar flares can severely impact the quality of atmospheric signals and therefore need to be properly studied, both to mitigate their effects and to exploit them as a diagnostic tool for star–planet interactions.
The analysis of stellar flares serves multiple purposes: it allows the identification and exclusion of problematic targets where frequent flaring could compromise transit observations; it provides insights into the planetary radiation environment, with direct implications for atmospheric escape and photochemistry; it enables the inclusion of magnetically active stars in dedicated investigations of star–planet magnetic interactions; and it supports the development of specific analysis techniques to model or remove flare signatures from light curves, ensuring more reliable atmospheric inference.\\
Building on the methodology originally developed by \cite{colombo2022short}, who introduced a flare-fitting framework applied to individual active stars such as DS Tuc and AU Mic, \cite{galletta2025exploring} extended this approach to a volume-limited sample of all M-type stars within 10 pc, enabling a systematic characterization of flare activity at the population level. In the present study, we further adopt and apply the same flare analysis framework to a different and substantially larger set of stellar targets selected from the Ariel target catalog, in order to investigate stellar activity across a broader range of spectral types and luminosities. This analysis is performed using TESS light curves \citep[][]{ricker2016transiting}.\\
The paper is structured as follows. Sect. \ref{sec:methods} describes the sample, the method used to analyze flare properties and the definition of the flare energy index. Sect. \ref{sec:results} shows the results obtained from the analysis. Sect. \ref{sec:Discussion} presents the discussion of the results in the context of previous works and their implications. In the end in Sect. \ref{sec:conclusions} summarizes our conclusions.
\section{Methods}\label{sec:methods}
In this section, we describe the methodology adopted to analyze the flare activity of the stellar sample. In Sect. \ref{sec:sample} we outline the selection criteria and characteristics of the sample. In Sect. \ref{sec:flareproperties} we detail the procedure used to identify individual flares and to measure their main physical properties, such as amplitude, duration, and total emitted energy. In Sect. \ref{sec:validation_test} is presented a validation test of the flare detection and characterization pipeline, used to assess completeness and potential systematic effects. Finally, in Sect. \ref{sec:flareindex}, we introduce the Flare Energy Index (GF.01), a quantitative parameter defined to characterize the overall flare activity of each star and to enable comparison across different stellar populations.
\subsection{The sample}\label{sec:sample}
\begin{figure}
    \centering
    \includegraphics[width=\hsize]{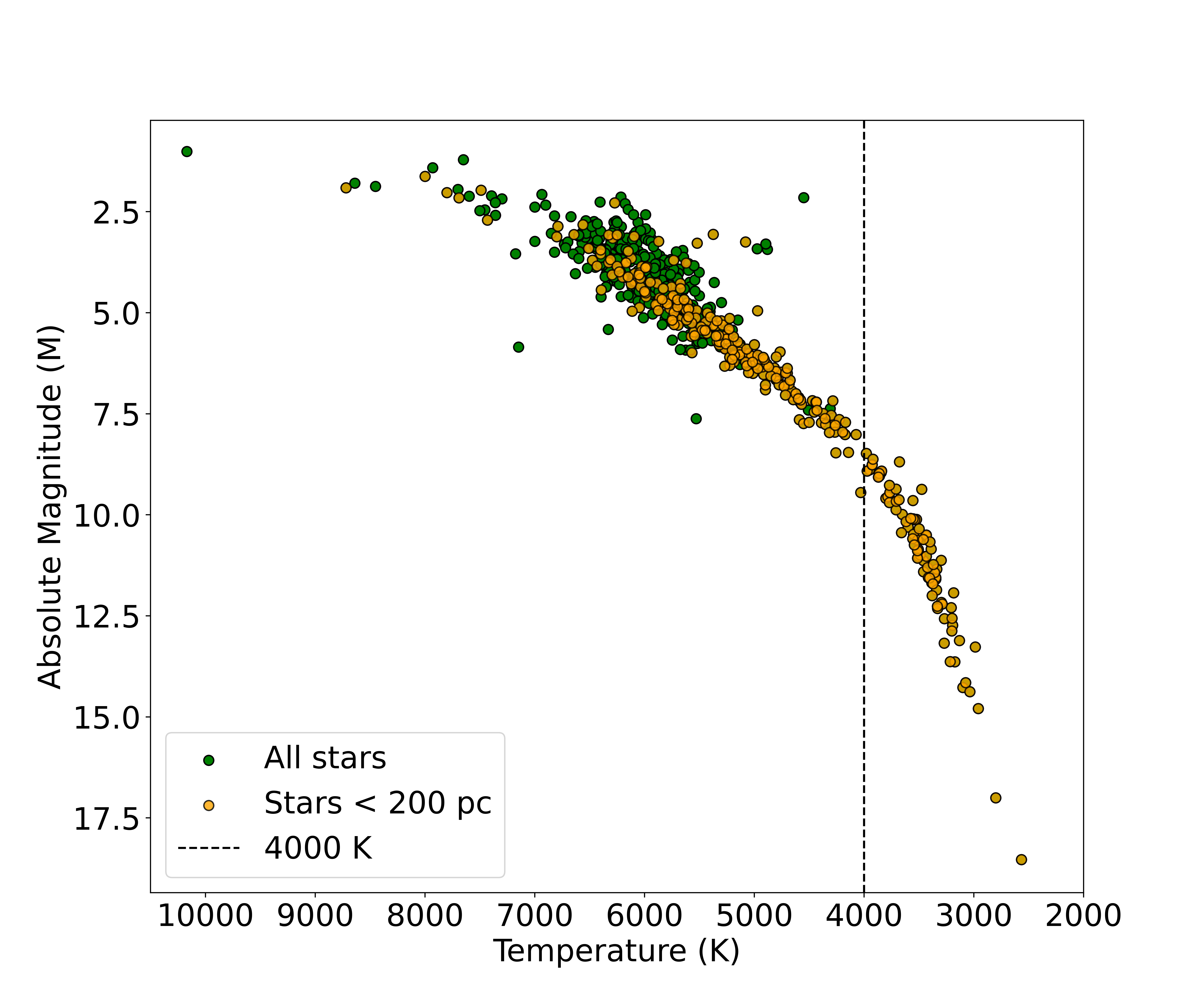}
    \caption{HR diagram of confirmed Ariel targets with TESS light curves (green), compared to our sample, which comprises all targets within 200 pc with available TESS light curves (290 targets total, orange). The comparison highlights that the 200 pc subsample spans the same temperature-luminosity range, ensuring representativeness of the full sample. The black dotted line represent the upper temperature limit of M dwarfs.}
    \label{fig:hrdiagram}
\end{figure}

\begin{figure}
    \centering
    \includegraphics[width=\hsize]{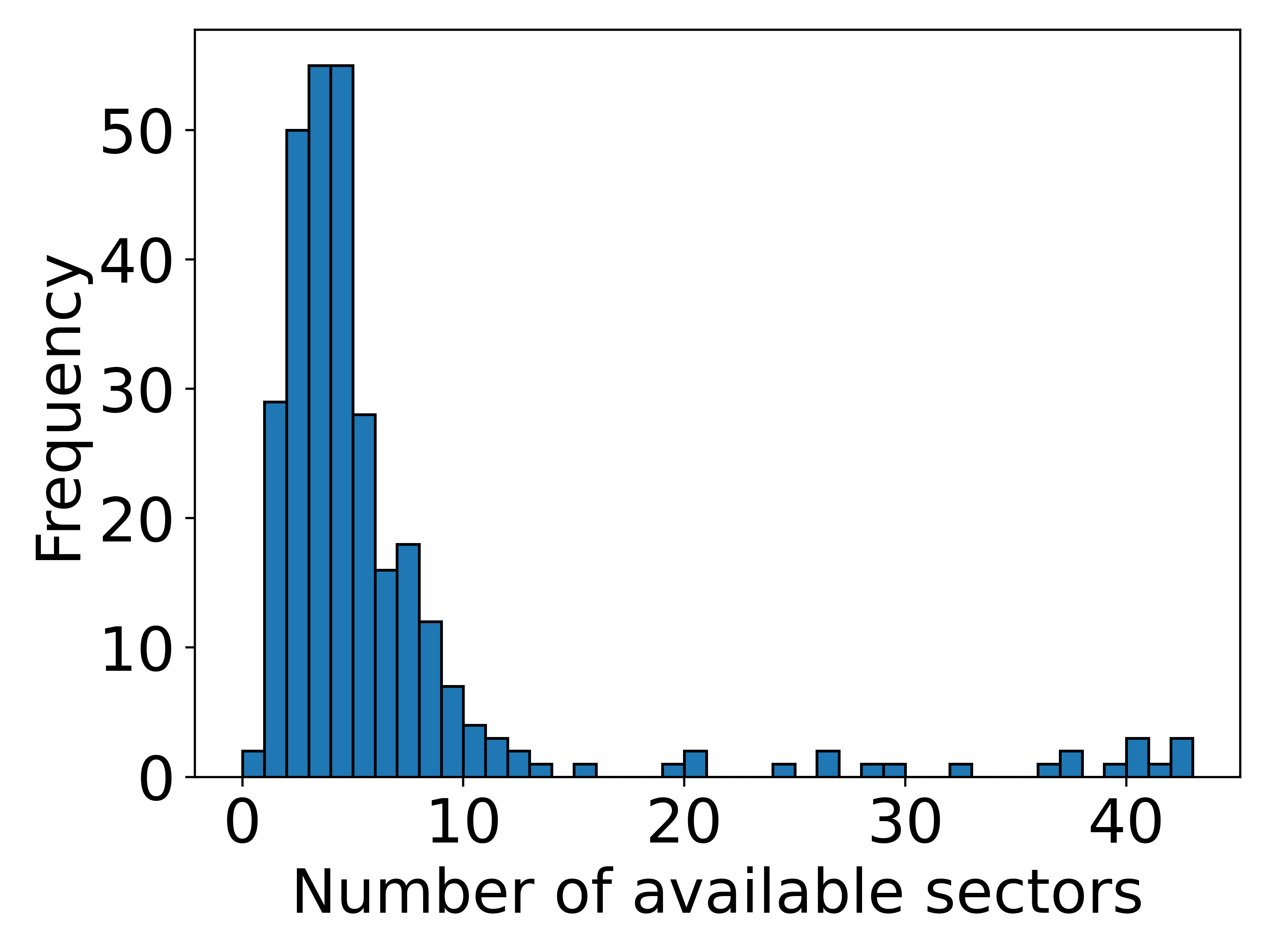}
    \caption{Number of TESS sectors available for the 290 targets in our sample.}
    \label{fig:istogramma}
\end{figure}

The current study is based on the Ariel Mission Candidate Sample (MCS) released in July 2024, which includes confirmed exoplanets only.
The Ariel MCS is based on the Ariel target list methodology described by \cite{edwards2022ARIEL} and developed using the ArielRad performance simulator \citep[][]{mugnai2020arielrad}. The 2024 version of the catalog also builds upon the procedures and selection criteria first introduced in \cite{edwards2019updated}, ensuring a consistent and optimized target selection for atmospheric characterization within the Ariel mission framework.
The 2024 MCS catalog includes:
\begin{itemize}
    \item 722 confirmed planets
    \item 657 stars (303 stars < 200 pc)
    \item 612 stars with available TESS light curves
    \item 290 stars with TESS light curves and distance < 200 pc
\end{itemize}
For this flare analysis, we focus on the subsample of 290 nearby stars.
This analysis is particularly relevant to mission planning: knowing the expected flare probability allows for estimating the number of visits required to reach a given signal-to-noise ratio in the presence of flares, as well as for identifying targets that may need specific data processing.\\
To contextualize the sample of stars within 200 pc used for the flare analysis, in Fig.\ref{fig:hrdiagram} we compare their distribution in the Hertzsprung-Russell (HR) diagram with that of the entire sample of 612 confirmed Ariel targets with available TESS light curves. Fig. \ref{fig:istogramma} shows the distribution of TESS sectors available per star. A larger number of observed sectors improves our ability to constrain the flare distribution and increases the likelihood of detecting rare, high-energy events. Most of the targets in our sample have 4 sectors, while only 20 are observed in more than 15 sectors.
The method used to analyze flares and distribution is based on \cite{galletta2025exploring}.

To retrieve the TESS data, we used the Lightkurve package \citep[][]{2018ascl.soft12013L} using as search parameters a cadence of 120 seconds and a search radius of 0.5 arcsec. Specifically, we
searched for light curves with 120s cadence data from the SPOC (Science Processing Operations Center) pipeline. The target columns retrieved for analysis were \textit{TIME} and \textit{SAP\_FLUX}, where \textit{SAP\_FLUX} refers to the Simple Aperture Photometry flux, i.e. the raw flux extracted by summing the calibrated pixel values within a predefined aperture, without correction for instrumental systematics or contamination from nearby sources. In addition, we apply the standard TESS quality flags and remove all cadences with non-zero quality indicators to exclude known instrumental artifacts and anomalous data points, following standard TESS data reduction practices.
\subsection{Measuring Flare Properties}\label{sec:flareproperties}
\begin{figure*}
    \centering
    \subfloat[]{
        \includegraphics[width=0.32\hsize]{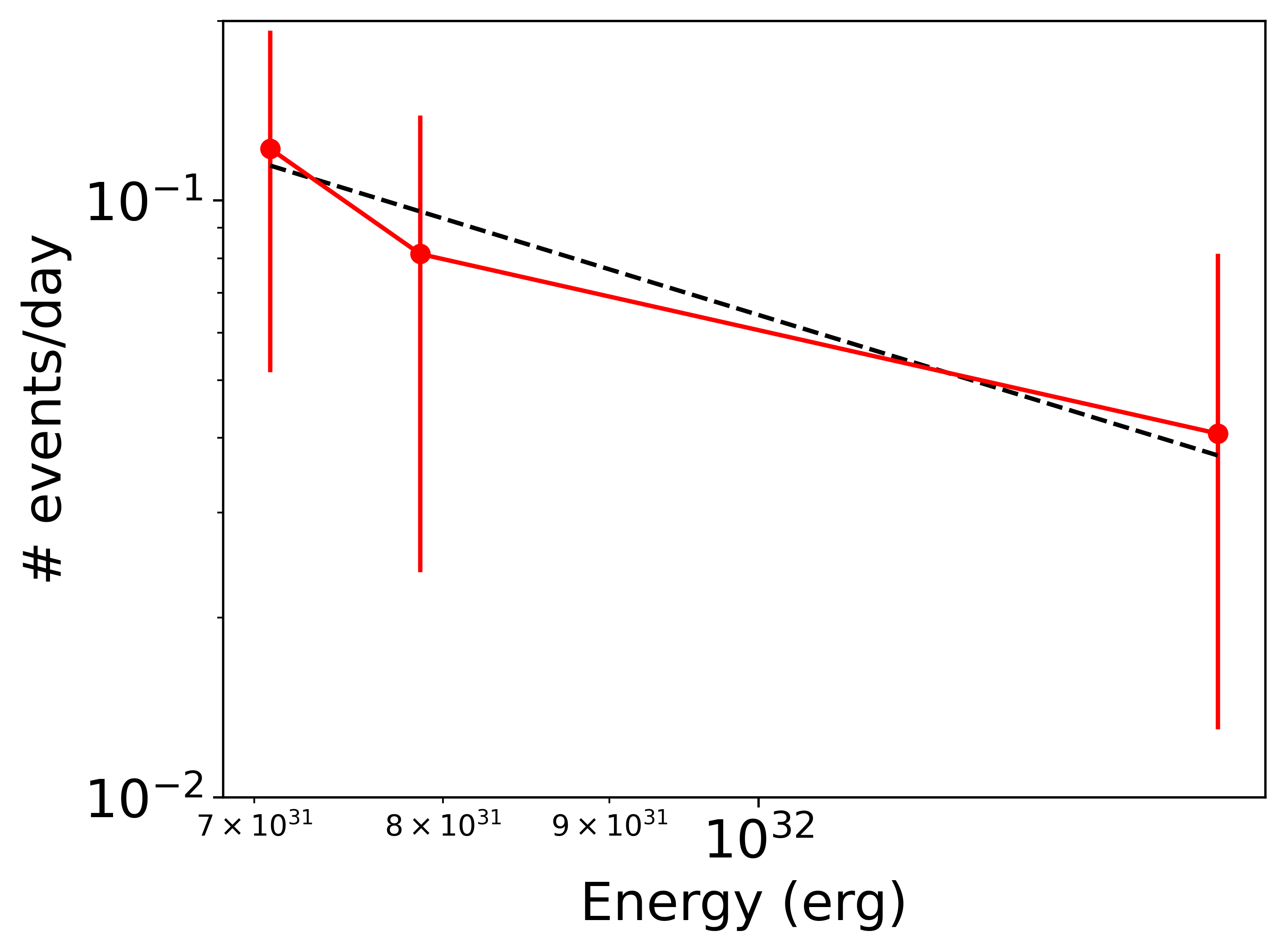}
        \label{fig:plot_one_segment_slope}
    }
    \subfloat[]{
        \includegraphics[width=0.32\hsize]{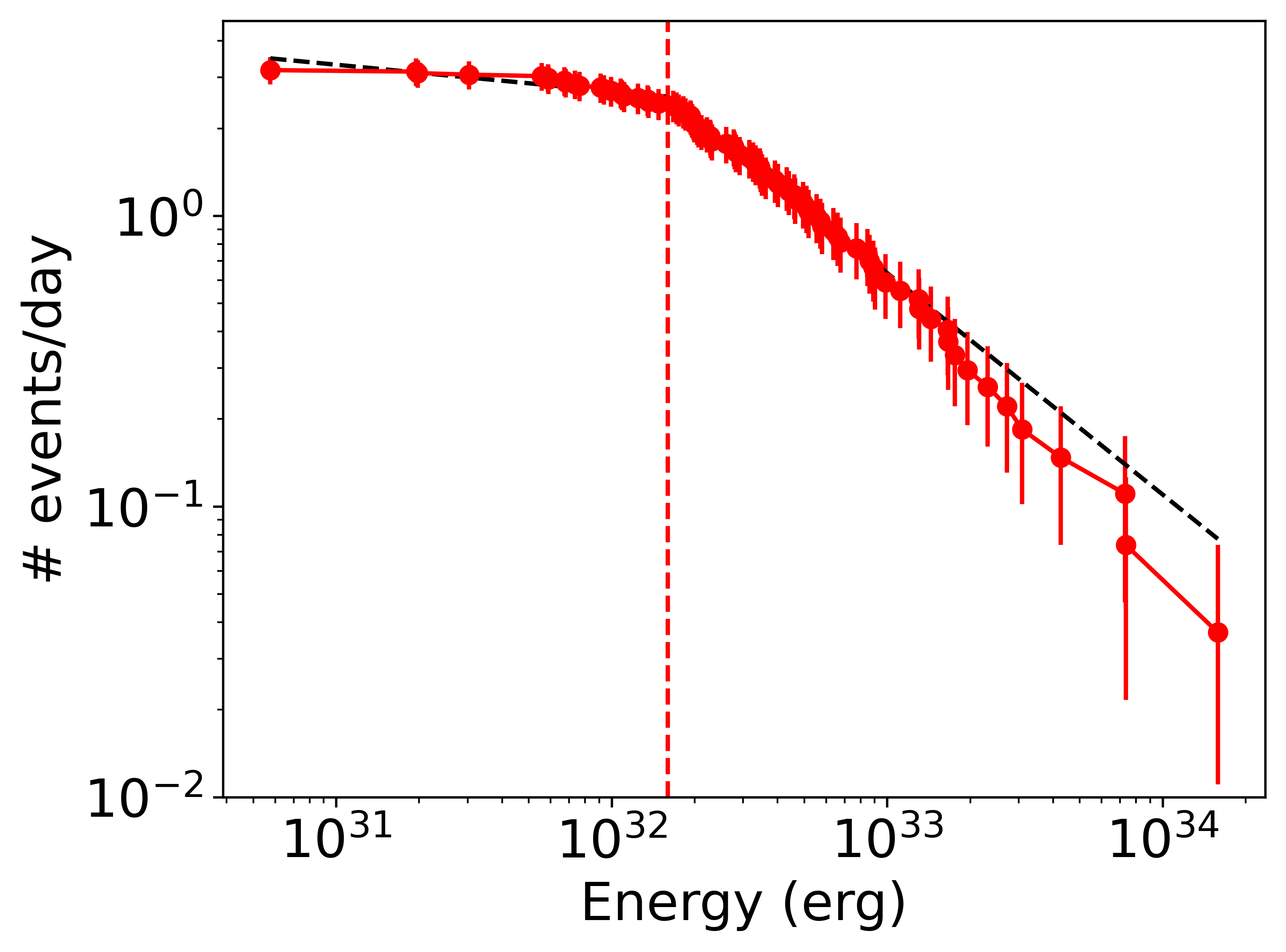}
        \label{fig:plot_two_segment_slope}
    }

    \caption{(a) Example of one-segment slope of a sector of WASP-189. (b) Example of two-segment slope of a sector of AU Mic. The red line with error bars is the observed cumulative curve of the star in a sector, the black dashed lines represent the fit and the red dashed vertical line (only in two-segment) represents the break point.}
    \label{fig:segment_slope_examples}
\end{figure*}

Following the method used in \cite{colombo2022short} and \cite{galletta2025exploring} we identify flares and then for each one, we derived the amplitudes, timescales, and total energy emitted.
To isolate flare signals, we first remove long-term stellar variability from the light curves. This is achieved using an iterative Gaussian Process (GP) regression with a SHOTerm kernel which is a term representing a stochastically-driven, damped harmonic oscillator \citep[][]{foreman2017fast}, which is well suited to describe correlated variability on timescales longer than typical flare durations. 
The SHOTerm kernel is parameterized by three hyperparameters (amplitude, characteristic period, and damping timescale), with the adopted priors defined in the range $\sigma \in [10^{-5},10^{5}]$, $P \in [1,50]$, and $\tau \in [10^{3},10^{4}]$, where all quantities are expressed in units of the light curve time axis (2-minute TESS cadences, corresponding to 120 s). These choices ensure that the Gaussian Process captures variability on timescales of $\sim 1$–$10$ days, while preserving short-duration flare signals, which typically occur on timescales of minutes to hours. The lower bound on the characteristic timescale is therefore well separated from the typical flare duration range adopted in this work, preventing the GP from absorbing impulsive flare-like events.\\
At each iteration, we compute a Maximum A Posteriori (MAP) GP model and subtract it from the data to obtain residuals. A 3$\sigma$ clipping procedure is then applied to remove outliers, and the GP is re-fitted to the cleaned data. This process is repeated until the number of outliers is consistent with statistical expectations.\\
The final GP model produced at the end of this iterative sigma-clipping procedure is therefore constrained to model only the long-term variability in the light curve. Since flare-like events are removed during the iterative sigma-clipping, the GP is not able to reproduce short impulsive signals, ensuring that flares are preserved in the final residual curve (RC), which is obtained by subtracting the final GP model from the original light curve.\\
Short-term events are identified in the RC as significant local maxima above a 3$\sigma$ threshold, computed from the statistical properties of the residuals. These events are characterized by a rapid rise, a slower decay, durations of minutes to hours, and peak fluxes exceeding the standard deviation of the residual light curve by at least $3\sigma$, allowing us to distinguish them from residual systematics and noise.\\
Each flare is modeled using a two-part exponential function, consisting of an increasing exponential for the rise phase and a decaying exponential for the decay phase, joined at the flare peak. This model captures flare morphology and provides robust estimates of amplitudes and characteristic timescales.\\
For each star, we construct the cumulative flare energy distribution, representing the number of flares above a given energy. These distributions are fitted using a two-segment power law in logarithmic space. The two segments are separated by an energy break corresponding to a change in slope.\\
We perform piecewise linear fitting in logarithmic space using a bootstrap resampling approach to estimate the uncertainties on both the slopes and the breakpoint energy. If fewer than six flares are detected for a given star, only a single power-law fit is applied, as the data do not allow a reliable determination of a breakpoint. This condition applies to 316 out of 1638 analyzed sectors. Additionally, in cases where fewer than two flares are present above the energy break (66 out of 1638 sectors), only the first slope is retained because we are unable to fit a slope to only one point. 
Fig. \ref{fig:segment_slope_examples} shows both example of the single- and the two-segment fits, respectively using sectors of WASP-189 and AU Mic, the black dashed lines represent the fit and the red dashed vertical line represents the break point.

\subsection{Validation test}\label{sec:validation_test}
\begin{figure*}
    \centering
    \subfloat[]{
        \includegraphics[width=0.32\hsize]{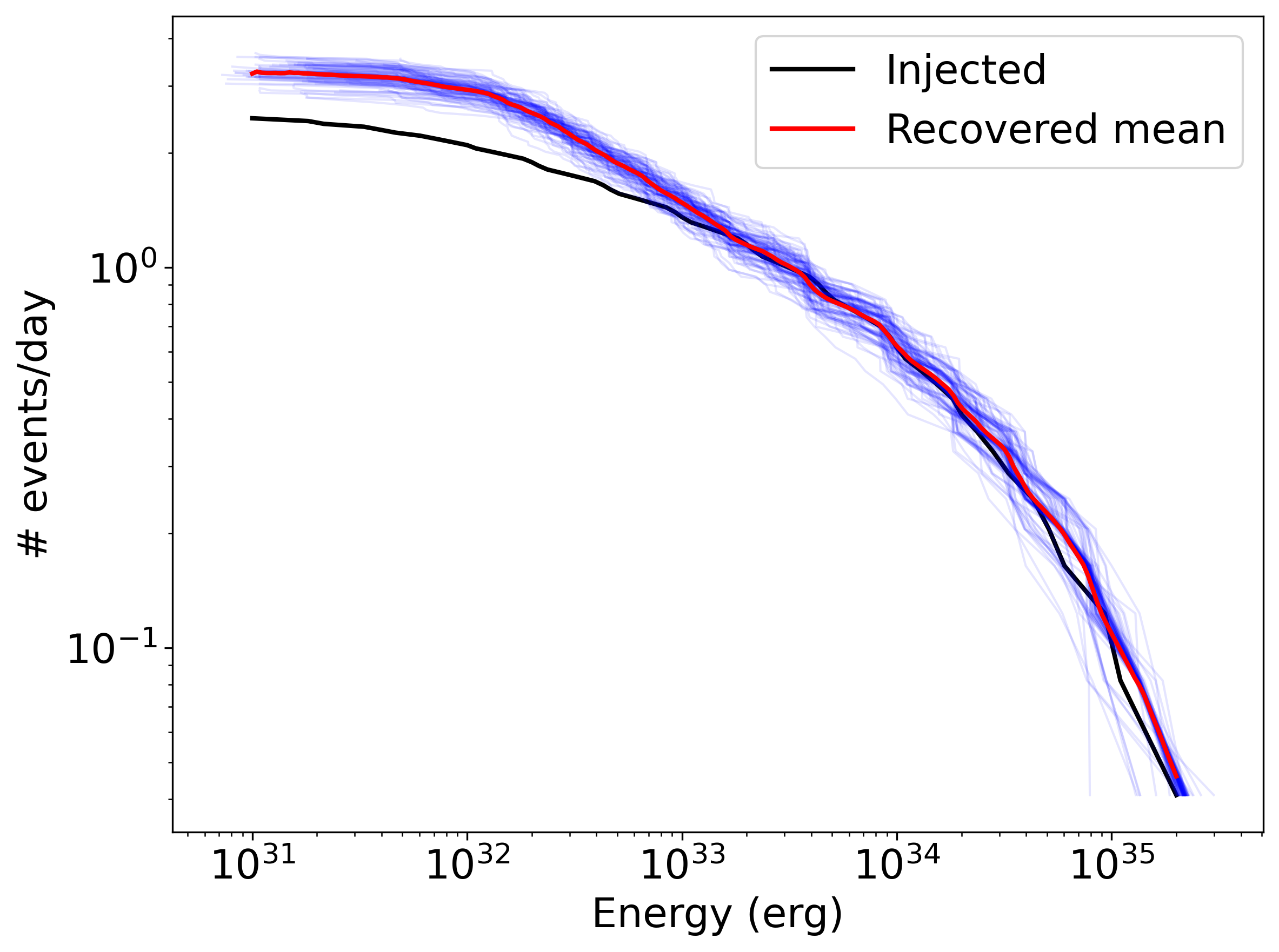}
        \label{fig:FFD_real_vs_injected}
    }
    \subfloat[]{
        \includegraphics[width=0.36\hsize]{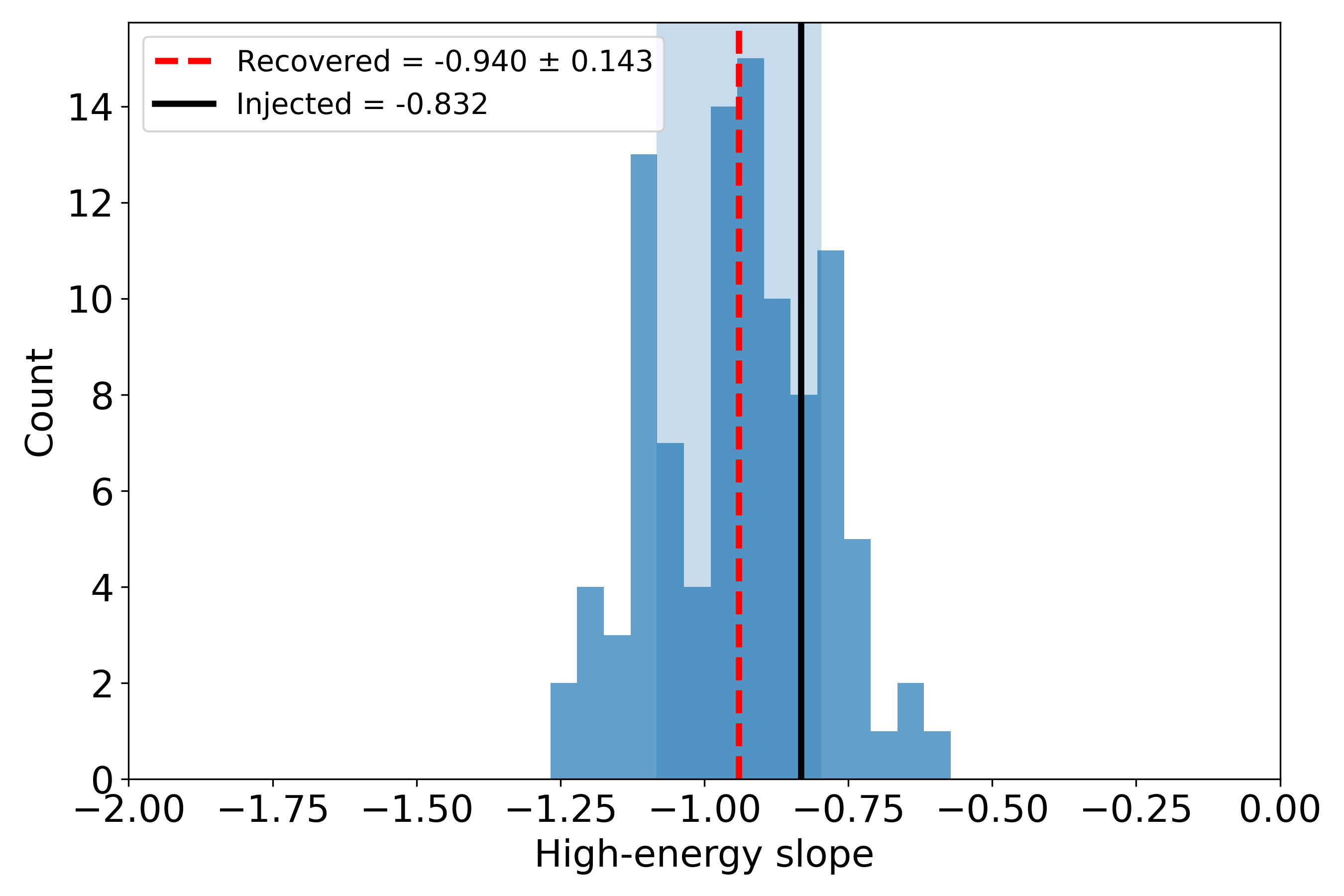}
        \label{fig:slope_distribution}
    }

    \caption{Injection--recovery validation tests of the flare detection pipeline. (a) Comparison between the FFD of injected and recovered synthetic flares. The injected FFD is shown as a black curve, while the recovered distributions from 100 realizations are displayed in blue. The red curve represents the mean recovered FFD. The comparison reveals that the recovered distribution shows an excess of low-energy events, whereas high-energy events are recovered with high efficiency. (b) Distribution of recovered power-law slopes obtained from 100 synthetic realizations with an input slope of $-0.832$. The vertical black line indicates the injected slope, while the red dashed line and shaded region show the mean recovered value and its error.}
    \label{fig:validation_test}
\end{figure*}

To evaluate the completeness of the flare detection pipeline and quantify possible biases in the inferred FFDs, we performed injection-recovery simulations using synthetic flare events. Artificial flares were injected into light curves constructed directly from the observed TESS photometry of AU Mic, preserving the original temporal sampling, data gaps, and instrumental systematics. This ensures that the simulations fully retain the intrinsic red-noise properties of the stellar variability as well as all observational effects present in the real dataset.\\
Synthetic flares were generated by randomly sampling flare amplitudes and characteristic timescales over broad parameter ranges representative of the observed flare population. In particular, flare energies were sampled over the interval $10^{31}$--$10^{35}$ erg using a logarithmically spaced grid, while flare durations were sampled between $\sim100$ and $\sim2000$ seconds. The injected flare model follows the same two-component exponential model adopted for the flare fitting procedure. The injected flare model is defined as a piecewise exponential function describing a rapid rise followed by a slower decay:
\begin{equation}\label{injection_equation}
    F(t) =
\begin{cases}
A e^{\left(\frac{t - t_0}{\tau_r}\right)}, & t \le t_0 \\
A e^{\left(-\frac{t - t_0}{\tau_d}\right)}, & t > t_0
\end{cases}
\end{equation}
where $A$ is the flare amplitude, $t_0$ is the peak time, $\tau_r$ is the rise timescale, and $\tau_d$ is the decay timescale. The model is evaluated within a finite temporal window of $\pm 5\tau$ around the flare peak to reproduce the typical observable duration of stellar flares in TESS light curves.\\
The rise and decay components are constrained such that $\tau_r : \tau_d = 1 : 10$, corresponding to a total timescale $\tau = \tau_r + \tau_d$. This choice ensures that the decay phase is systematically longer than the rise phase ($\tau_d > \tau_r$), enforcing physically realistic flare asymmetry consistent with observed stellar flare profiles. \\
A total of 60 synthetic flare events were injected at random times within the valid observing windows and analyzed using the same reduction and detection pipeline adopted throughout this work.\\
Across the 100 bootstrap realizations, the number of recovered flares ranges from 68 to 89, with a mean value of $\sim$79 recovered events out of the 60 injected flares. Fig. \ref{fig:FFD_real_vs_injected} shows the comparison between the FFDs of the injected and recovered samples. The injected population is shown as a black curve, while the recovered distributions from individual realizations are displayed in blue, and their mean trend is shown in red. We can see that the recovered FFD exhibits an excess of low-energy events. A visual inspection of the detected flares indicates that this excess is mainly produced by noise spikes that occasionally satisfy the adopted $3\sigma$ detection criterion and are therefore classified as flare candidates. At higher energies, however, the recovered FFD closely follows the injected distribution, demonstrating that the detection pipeline reliably reproduces the intrinsic flare population in the regime where the signal-to-noise ratio is sufficiently high. This behavior further justifies the choice adopted throughout this work of restricting the analysis to the high-energy portion of the FFD.\\
Fig. \ref{fig:slope_distribution} shows the distribution of recovered power-law slopes obtained from 100 synthetic flare populations with an injected slope of $-0.832$. The distribution peaks close to the injected value, with a mean recovered slope of $-0.940 \pm 0.143$. This indicates that the adopted methodology does not introduce significant systematic biases in the determination of the high-energy slope and that the observed offset remains small compared with the statistical uncertainties.\\
For this reason, throughout the analysis we adopt a two-segment power-law model and derive the GF.01 index (see Sect. \ref{sec:flareindex}) exclusively from the high-energy slope. The injection--recovery simulations show that this portion of the FFD provides the most reliable characterization of the flare population, whereas the low-energy regime is more strongly affected by contamination from spurious detections generated by noise fluctuations near the detection threshold.
\subsection{The Flare Energy Index (GF.01)}\label{sec:flareindex}
To quantify flare activity in a normalized way, we adopt the GF.01 index, introduced in \cite{galletta2025exploring}, which quantifies the flare activity normalized to the stellar bolometric luminosity. The index represents the energy level at which the flare frequency equals 0.1 flares per day. This index is derived from the fitted power-law parameters (slope and intercept) and the star’s bolometric luminosity. GF.01 allows comparison across stars of different luminosities by normalizing the flare frequency–energy relation.
The GF.01 is given by the following equation:
\begin{equation}
    GF.01 = \frac{0.1-q}{m}/L_{bol}
\end{equation}
where $m$ is the slope of the high-energy segment of the two-segment flare frequency distribution (i.e. the portion unaffected by biases at low energies), $q$ is the intercept,
$L_{bol}$ is the bolometric luminosity of the star (we used the Virtual Observatory SED Analyzer (VOSA) to compute the bolometric luminosity of the stars \citep[][]{rodrigo2020vosa}), and 0.1 indicates the characteristic energy of flares occurring on average every 0.1 days.\\
GF.01 is computed independently for each sector of a given star, rather than on a star-averaged flare distribution. As a result, multiple GF.01 values may be associated with the same star. We compute GF.01 only for sectors with well-constrained two-segment fits. Sectors in which fewer than six flares are detected, and for which only a single power-law fit is available, are excluded from the analysis, as the absence of a well-defined high-energy slope prevents a reliable determination of the index. The use of the high-energy slope ensures that the derived values are not biased, which predominantly affects the low-energy regime of the flare frequency distribution (see Sect. \ref{sec:validation_test}).\\
Uncertainties on GF.01 are estimated through standard error propagation from the uncertainties on the fitted parameters. Assuming uncertainties $\sigma_m$, $\sigma_q$, and $\sigma_{L_{bol}}$, the uncertainty on GF.01 is given by:
\begin{equation}
\sigma_{GF.01} = GF.01 \cdot \sqrt{
\left( \frac{\sigma_q}{0.1 - q} \right)^2 +
\left( \frac{\sigma_m}{m} \right)^2 +
\left( \frac{\sigma_{L_{bol}}}{L_{bol}} \right)^2
}
\end{equation}
where the uncertainties on $m$ and $q$ are derived from bootstrap resampling of the flare frequency distribution fits.

This normalization accounts for differences in stellar luminosity, thereby enabling an assessment of how efficiently a star converts its available energy into flare energy. A higher GF.01 value indicates that, on average, the flares are more energetic, whereas a lower GF.01 value corresponds to less energetic flares.
\section{Results}\label{sec:results}
In this section, we present the main results of our analysis. Overall, our flare-search pipeline identified a total of 15857 flare events across 290 analyzed stars and 1638 TESS sectors. The number of detected flares per sector ranges from a minimum of 2 up to a maximum of 86 events.
In Sect.~\ref{sec:slopeanalysis} we report the outcomes of the slope analysis, which provides information on the shape and energy distribution of the detected flares. In Sect.~\ref{sec:indexanalysis} we discuss the behavior of GF.01 across the stellar sample and its dependence on stellar parameters. In Sect.~\ref{sec:probability} we evaluate the probability that stellar flares occur during planetary transits, using the FFDs to estimate the expected occurrence of energetic events during Ariel observations and to assess their potential impact on exoplanet detection and characterization. 
\subsection{Slope analysis}\label{sec:slopeanalysis}
\begin{figure*}
    \centering
    \subfloat[]{
        \includegraphics[width=0.32\hsize]{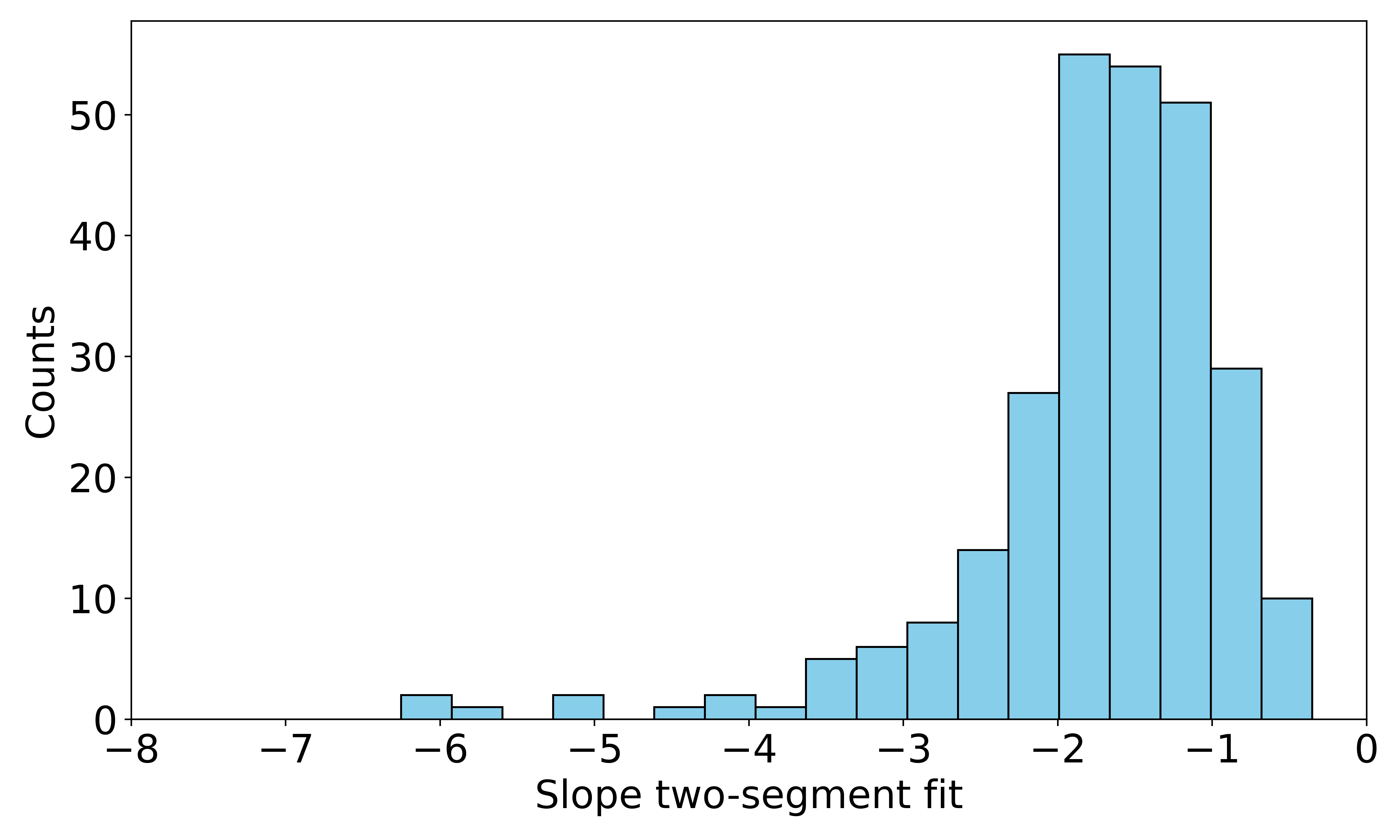}
        \label{fig:hist_slope_second_mean}
    }
    \subfloat[]{
        \includegraphics[width=0.32\hsize]{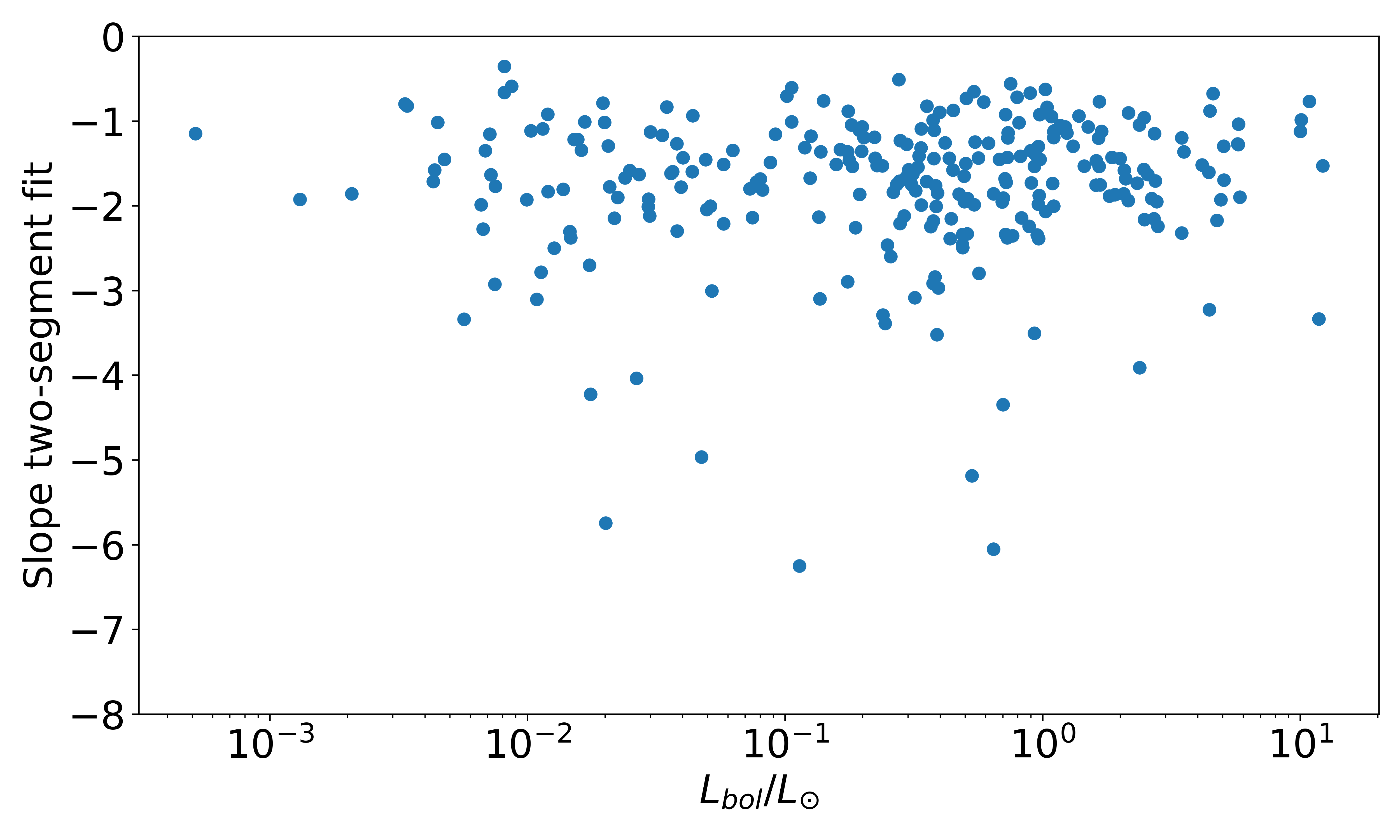}
        \label{fig:slope_second_mean_vs_Lbol_mean}
    }
    \subfloat[]{
        \includegraphics[width=0.32\hsize]{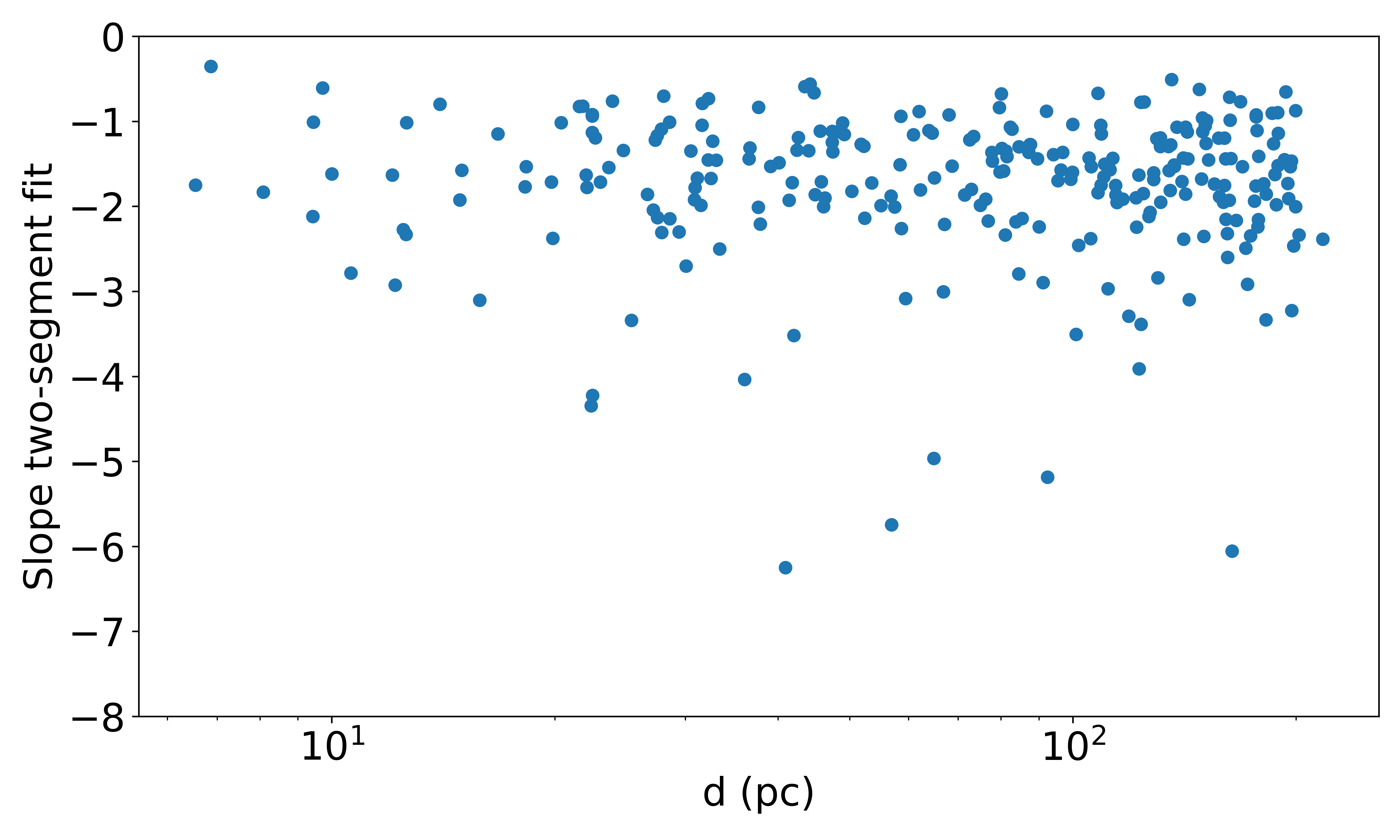}
        \label{fig:slope_second_mean_vs_d_mean}
    }
    \caption{(a) Distribution of the high-energy slopes derived from the two-segment power-law fits of the flare frequency distributions (FFDs). The distribution shows a clear peak around $-1.8$, with a mean value of $-1.87$, a median of $-1.59$, and a standard deviation of $1.55$, indicating a broad but well-defined range of flare activity across the sample.
    (b) High-energy slope vs stellar bolometric luminosity ($L_{\mathrm{bol}}$). No clear correlation is observed, although more luminous stars tend to show a reduced dispersion in slope values, suggesting a more homogeneous flare energy distribution in this regime.
    (c) High-energy slope vs stellar distance. No significant trend is detected across the explored distance range, indicating that the derived slopes are not strongly affected by distance-dependent observational biases.}
    \label{fig:segments}
\end{figure*}

The FFDs of the analyzed stars are constructed as cumulative occurrence rates of flares above a given energy and are modelled with a two-segment power-law fit following the methodology described in \cite{galletta2025exploring}. The two-segment model consists of two power laws joined at a break energy, and it is introduced to separate the energy range where observational biases dominate from the range where the flare occurrence can be robustly characterized. \\
At low energies, the observed distribution is strongly affected by detection biases: smaller flares have lower amplitudes and are progressively more difficult to detect above the photometric noise floor, leading to a flattening of the measured FFD. In this regime, the low-energy slope derived from the fit does not necessarily represent the intrinsic flare production physics of the star. This first segment, therefore, serves mainly to identify the onset of biases at low energies.
In contrast, the high-energy segment corresponds to the part of the distribution where the sample is more complete with respect to detection sensitivity, flares above the break energy are reliably detected across the full dataset. The high-energy slope of the FFD therefore provides a more robust characterization of the intrinsic flare occurrence rate, and can be interpreted in terms of the underlying magnetic energy-release processes. A steeper high-energy slope indicates that energetic flares are relatively rare compared to lower-energy ones, whereas a shallower slope implies a relatively higher incidence of more energetic flares.\\
The high-energy slope is particularly relevant because it determines the occurrence rate of energetic flares, which are the events most likely to be detectable and to significantly affect planetary transit observations. Energetic flares produce large and rapid increases in stellar flux, often comparable to or exceeding the transit depth, and can therefore alter key transit parameters such as the measured transit depth, duration, and ingress/egress shape if they occur during a transit.
Because the FFD quantifies the rate of flares above a given energy threshold, the high-energy power-law index allows us to estimate the expected number of flares with energies $E \geq E_{threshold}$ occurring within a finite time window, such as the duration of a planetary transit.
If the slope is flatter, energetic events are relatively more frequent, increasing the likelihood that a flare will coincide with an Ariel observation and contaminate the light curve. 
In contrast, steeper slopes indicate that high-energy events are rare, lowering the probability of strong flare contamination but still allowing moderate activity to affect transit depth measurements. \\
Figure \ref{fig:segments} presents the results of the two-segment power-law fit to the FFDs of the analyzed stars.
Figure \ref{fig:hist_slope_second_mean} shows the distribution of the high-energy slopes derived from the fits, which displays a clear peak around a value of approximately -1.8, representing the most common behavior within the sample. The mean high-energy slope is -1.87, with a standard deviation of 1.55 and a median value of -1.59.
These values are broadly consistent with previous studies of stellar FFDs, which have shown that flare energies follow power-law statistics over several orders of magnitude \citep[e.g.][]{aschwanden2021self, hawley2014kepler}. In particular, large-scale analyses based on Kepler data, such as the Kepler Catalog of Stellar Flares \citep[][]{davenport2016kepler}, demonstrate that stellar flare populations are well described by power laws, although the inferred slopes depend on stellar type, activity level, and completeness effects.
Studies of M dwarfs, including \cite{hawley2014kepler}, show that FFDs are consistent with power-law behavior at high energies, with deviations at low energies primarily due to observational biases. More recent analyses of TESS data, such as \cite{rajpurohit2025exploring}, find cumulative FFD slopes corresponding to high-energy slopes close to $-2$, in agreement with previous Kepler results.
Additional work on active M dwarfs in large TESS samples \citep[e.g.][]{capistrant2026stellar} confirms that typical slopes lie in the range $-1.5$ to $-2.0$, although significant star-to-star scatter is observed due to intrinsic variability and detection biases.
From a theoretical perspective, such power-law distributions are naturally explained by self-organized criticality models of magnetic reconnection, which predict scale-free flare energy release processes \citep[][]{aschwanden2021self}.
Overall, the average slope derived in this work ($-1.87$) lies well within the range reported in the literature and is consistent with results obtained for both individual active stars and large stellar samples, including those presented in \cite{galletta2025exploring}. This agreement indicates that the flare energy distributions in the analyzed sample follow the same statistical behavior observed in previous studies, supporting the robustness of the adopted detection and fitting methodology.\\
Figure \ref{fig:slope_second_mean_vs_Lbol_mean} shows the high-energy slope as a function of the stellar bolometric luminosity ($L_\mathrm{bol}$). No clear correlation between the two quantities is observed. However, stars with higher bolometric luminosities tend to exhibit a smaller dispersion in slope values compared to less luminous stars, suggesting a more homogeneous flare energy distribution within this subset of the sample.
Finally, Figure \ref{fig:slope_second_mean_vs_d_mean} presents the high-energy slope as a function of stellar distance. No significant trend is observed across the explored distance range. Any apparent variation is likely dominated by observational effects, such as changes in detection sensitivity with distance, rather than reflecting an intrinsic physical dependence.

\subsection{GF.01 analysis}\label{sec:indexanalysis}
\begin{figure*}
    \centering
    \subfloat[]{
        \includegraphics[width=0.32\hsize]{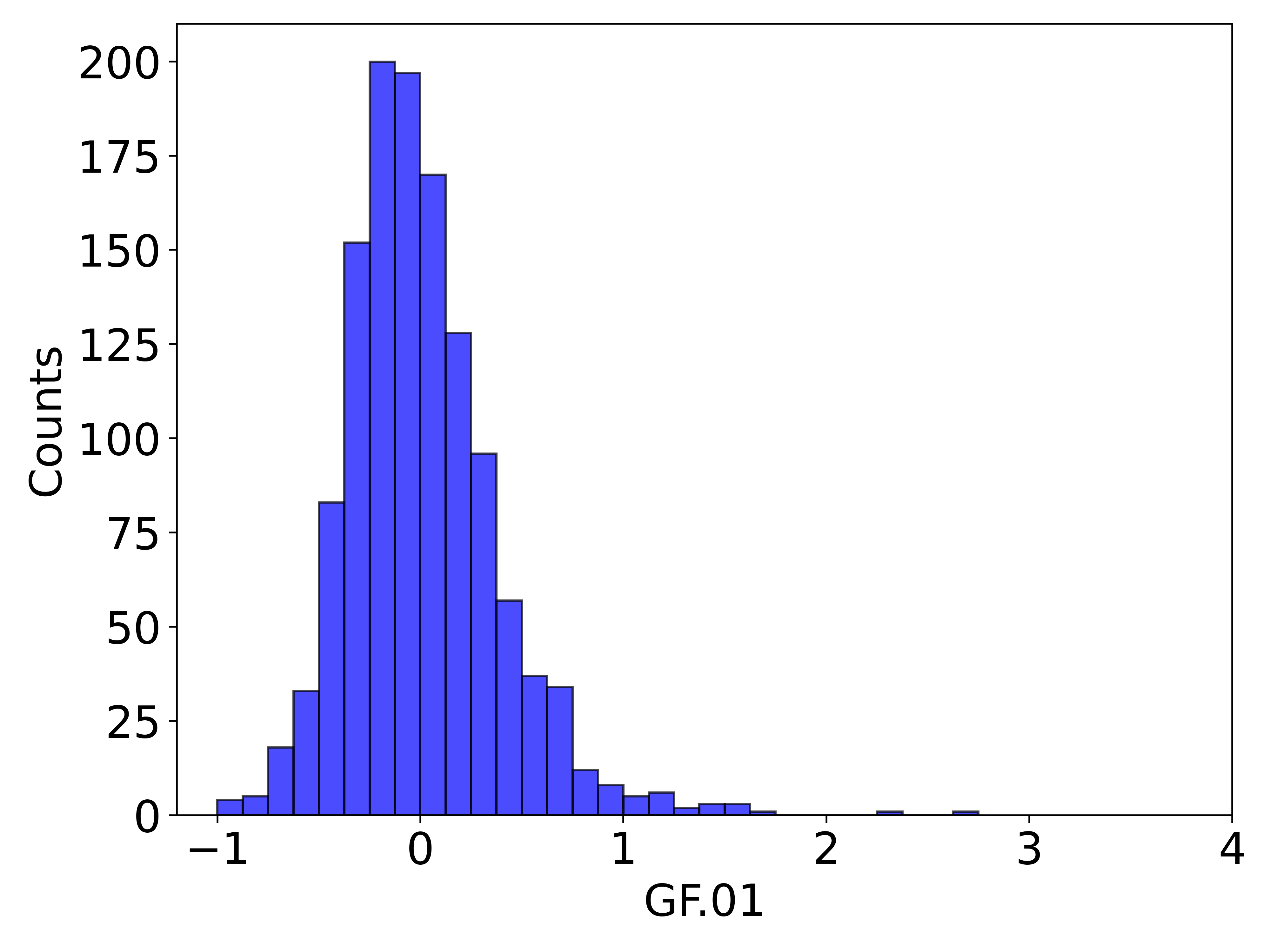}
        \label{fig:hist_gindex_ARIEL_0_200}
    }
    \subfloat[]{
        \includegraphics[width=0.32\hsize]{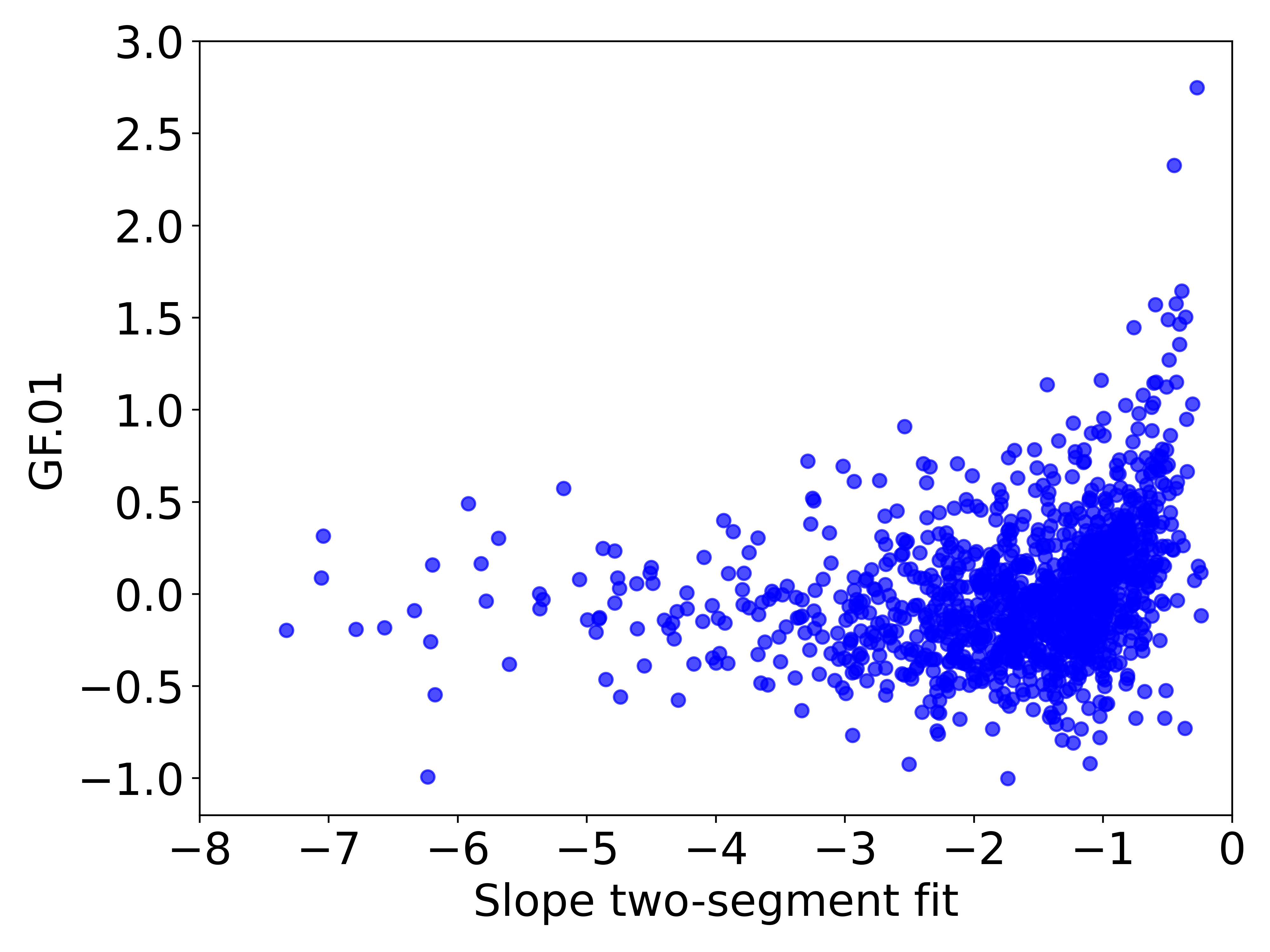}
        \label{fig:slopevsgindex_ARIEL_0_200}
    }
        \subfloat[]{
        \includegraphics[width=0.32\hsize]{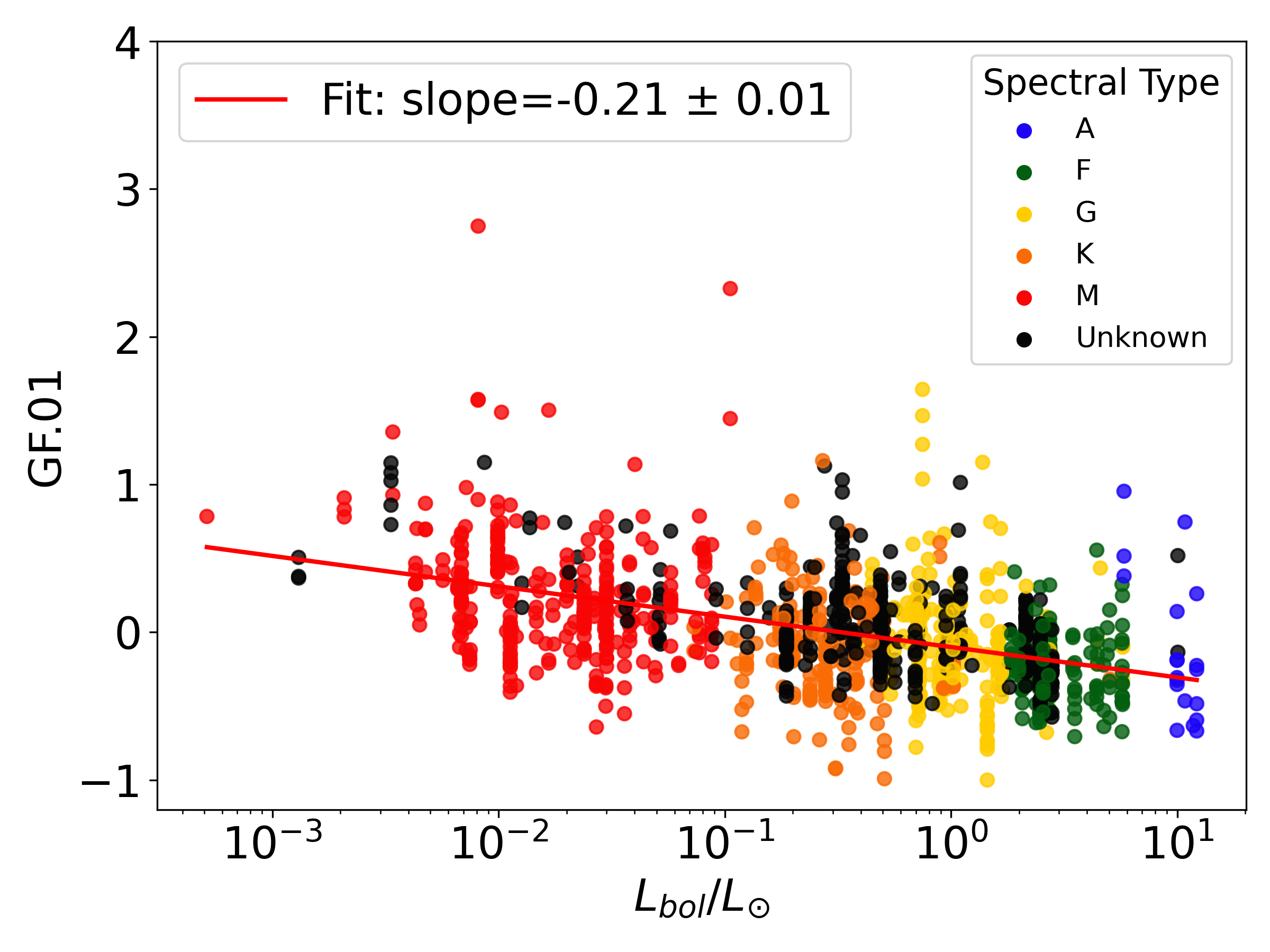}
        \label{fig:energyvsgindex_ARIEL_0_200}
    }
    \caption{(a) Distribution of the GF.01 index, which represents the flare energy corresponding to a frequency of 0.1 flares per day, normalized to the stellar bolometric luminosity. (b) GF.01 vs the high-energy slope of the two-segment flare frequency distribution, illustrating the relationship between flare energetics and the power-law behavior of the distribution. (c) GF.01 versus stellar bolometric luminosity ($L_\mathrm{bol}$), showing a clear anti-correlation between flare activity and stellar energy output. The red solid line indicates the best-fit linear relation in logarithmic space. The points are color-coded according to their spectral type classification (A, F, G, K, M), while black points correspond to sources for which no reliable spectral type could be retrieved.}
    \label{fig:gindex_ARIEL_0_200}
\end{figure*}

The GF.01 allows us to identify which stars are intrinsically more flare-active relative to their energy budget, making it a valuable diagnostic for evaluating the potential impact of flares on planetary environments and on the correct measurements of the Ariel observations.
By studying the distribution and correlations of GF.01, we can compare the activity of the Ariel sample with the sample of nearby M dwarfs and if their different behaviors is linked to selection effects, stellar type, or distance. 
In particular, combining GF.01 with the slopes of the FFDs provides insight into the relative occurrence of high-energy events, while the comparison with $L_{bol}$ tests whether the scaling of activity with luminosity remains consistent across diverse stellar spectral types. 
This framework is useful to estimate the probability of flare contamination during planetary transits and to better characterize the stellar environments probed by Ariel. \\
Fig. \ref{fig:hist_gindex_ARIEL_0_200} shows the overall distribution of GF.01, which displays a clear peak around zero, indicating that most stars exhibit relatively low or moderate flare activity compared to their bolometric output. A high-value tail is also visible, highlighting the presence of a small subset of particularly active stars.\\
Fig. \ref{fig:slopevsgindex_ARIEL_0_200} presents the comparison between GF.01 and the high-energy power-law slope derived from the two-segment model. Each data point corresponds to a single TESS sector rather than to an individual star, since both the slope and GF.01 are computed independently for each sector. The majority of data points cluster around GF.01 $\approx$ 0 and slope values between –2 and –1, suggesting that stars with intermediate flare activity tend to exhibit steep slopes, corresponding to a rapid decline in flare frequency with increasing energy.\\
Finally, Fig. \ref{fig:energyvsgindex_ARIEL_0_200} shows the relation between GF.01 and stellar bolometric luminosity ($L_\mathrm{bol}$), the points are color-coded according to their spectral type classification (A, F, G, K, M), while black points correspond to sources for which no reliable spectral type could be retrieved, the red solid line represents the fitted model. 
A clear negative relation emerges, confirming that less luminous stars display higher relative flare activity.
To quantify this behavior, we performed a linear fit in logarithmic space, obtaining a slope of $m = -0.206 \pm 0.011$ and an intercept of $q = -0.100 \pm 0.011$. The fit yields a $\chi^2$ value of $0.110$, indicating a statistically consistent description of the observed trend within the data uncertainties.
It is worth noting that GF.01 is computed using only high-energy flares, for which the detection completeness is significantly improved. This reduces the impact of observational biases related to flare detectability, ensuring that the derived trend is primarily driven by intrinsically robust flare events.
This behavior aligns with the findings of \cite{galletta2025exploring} for M-dwarf stars, where a similar trend was interpreted as evidence that magnetic activity represents a more significant fraction of the total energy output in lower-luminosity stars.

\subsection{Flare probability}\label{sec:probability}
\begin{figure}
    \centering

    \includegraphics[width=\hsize]{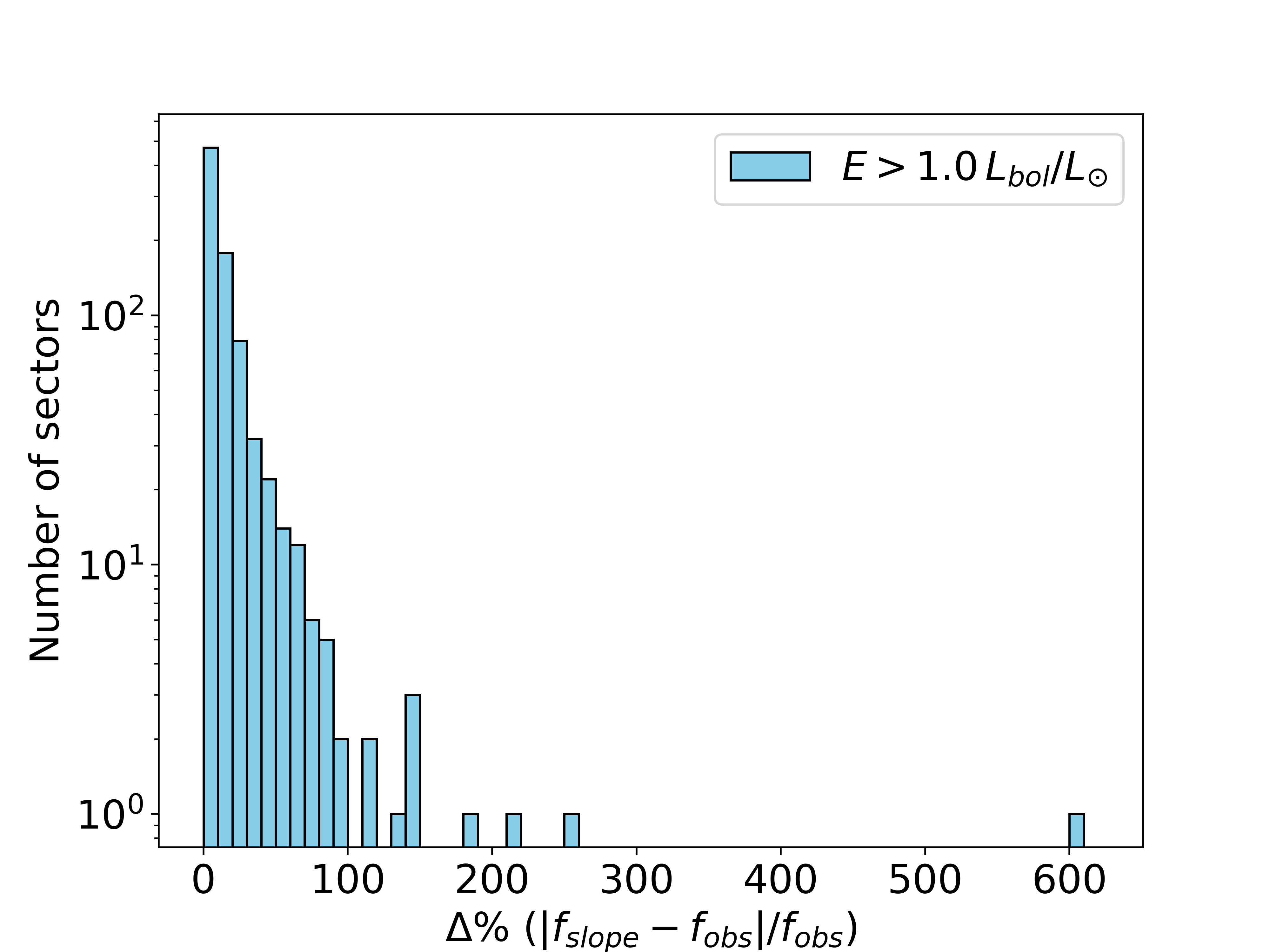}
    \caption{Distribution of the percentage difference $\Delta\%$ between the flare occurrence rates predicted from the fitted FFDs and those directly measured from the cumulative observed flare distributions, evaluated at an energy threshold of $1.0,L_{\mathrm{bol}}$. The quantity $\Delta\%$ quantifies the agreement between the model-based slope extrapolation and the empirical flare rate in each TESS sector, with larger values indicating sectors where the FFD fit significantly over or underestimates the observed flare frequency. The distribution shows that most sectors are in good agreement with 12\% (103/831 sectors) exceeding the 30\% quality cut applied in the analysis.}
    \label{fig:hist_delta_percent_sector_1.0}
\end{figure}

\begin{figure}
    \centering

    \includegraphics[width=\hsize]{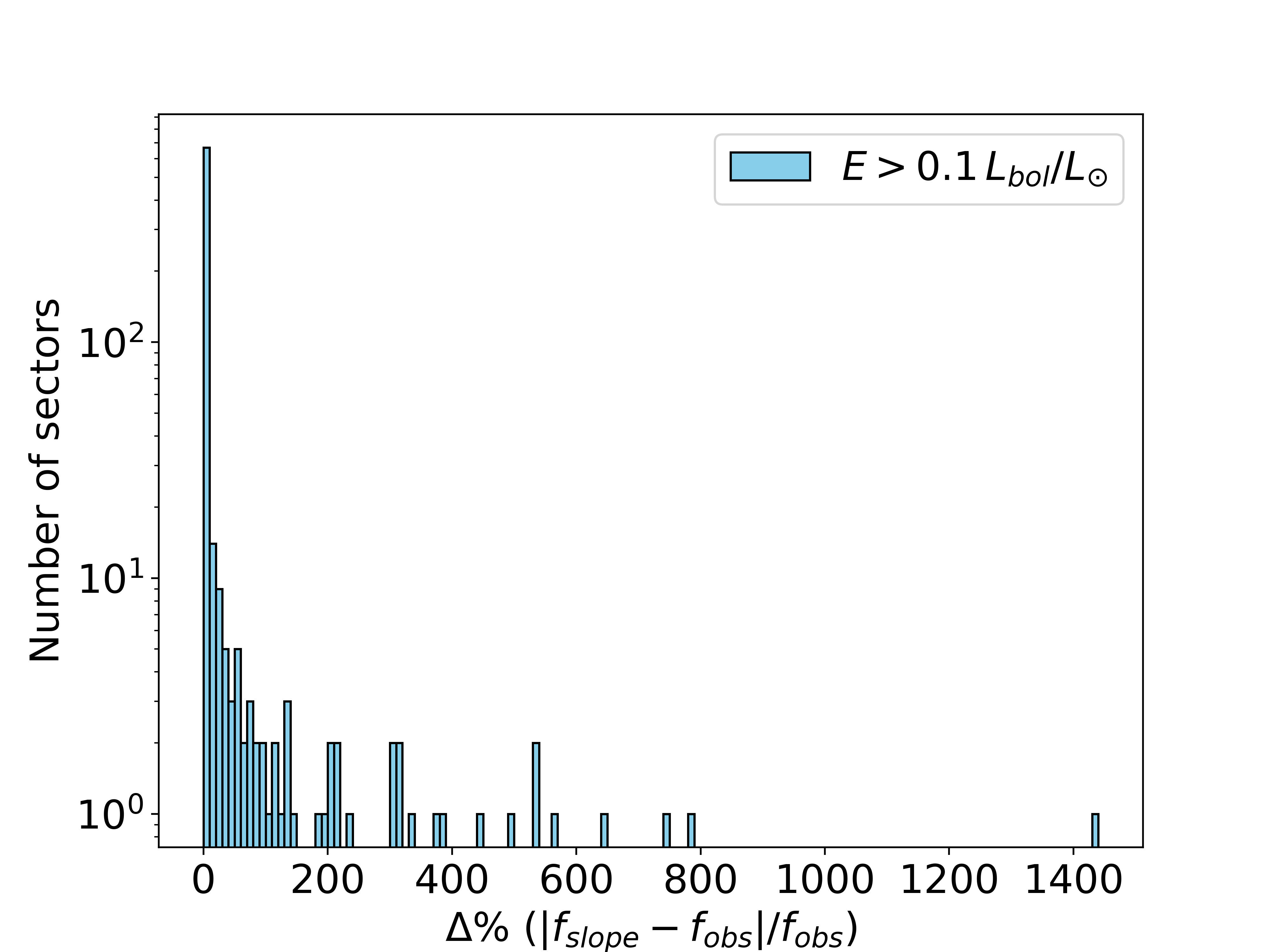}
    \caption{Distribution of the percentage difference $\Delta\%$ between flare occurrence rates predicted from the fitted FFDs and those directly measured from the cumulative observed flare distributions, evaluated at an energy threshold of $0.1,L_{\mathrm{bol}}$. The metric $\Delta\%$ provides a direct sector-by-sector diagnostic of the reliability of the FFD-based extrapolation used in the flare probability framework. As in the higher-threshold case, most sectors show good agreement between predicted and observed flare rates, with 7\% (53/744 sectors) exceeding the 30\% quality cut applied in the analysis.}
    \label{fig:hist_delta_percent_sector_0.1}
\end{figure}

\begin{figure}
    \centering

    \includegraphics[width=\hsize]{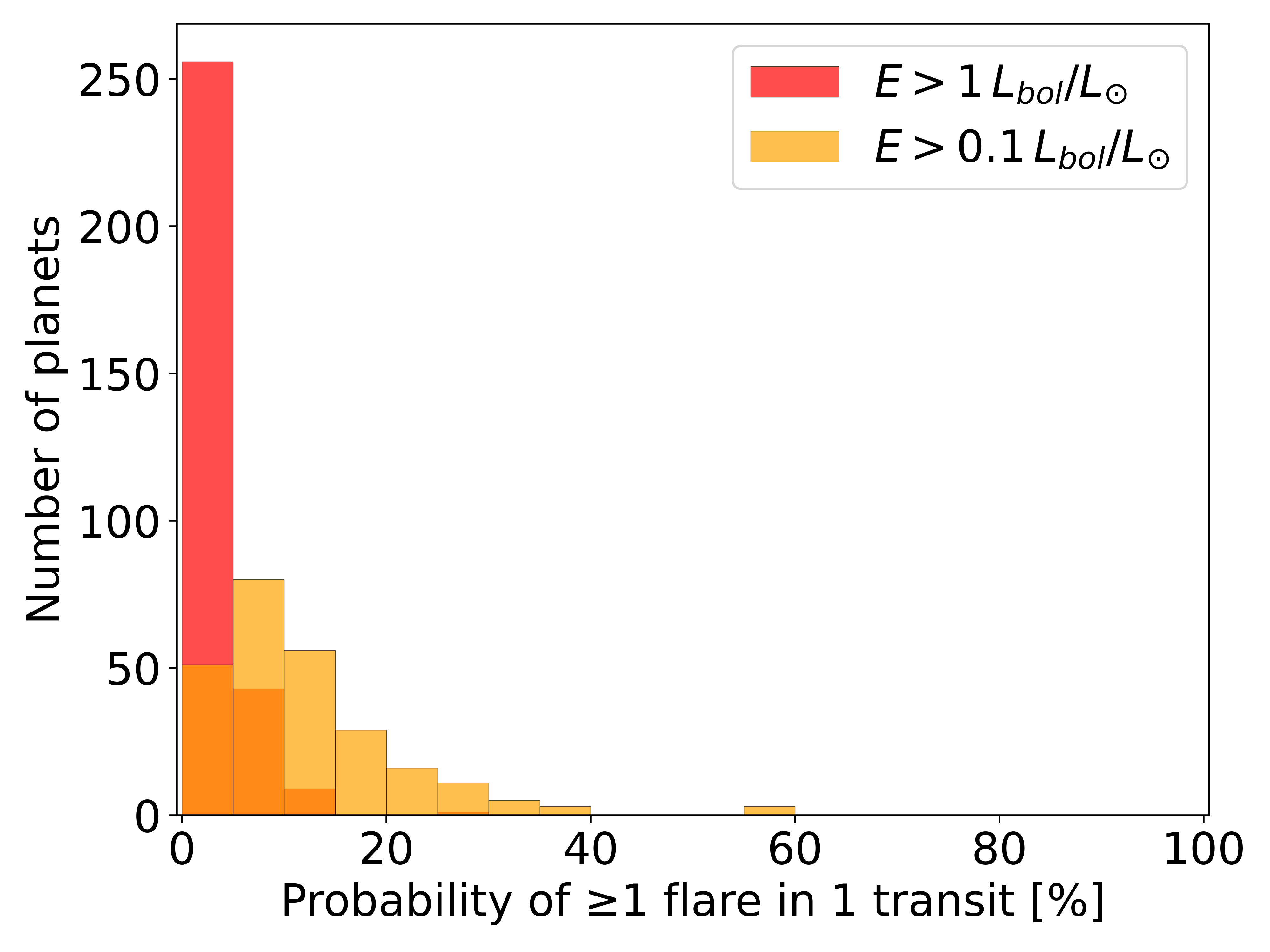}
    \caption{Distribution of flare probability during Ariel visit. The orange are the stars with the probability to have flare of the 10\% of energy of the $L_{bol}$ of the star, the red are the stars with the probability to have flare of the 100\% of energy of the $L_{bol}$}
    \label{fig:flare_probability}
\end{figure}

The primary goal of this analysis is to quantify the likelihood that an Ariel transit observation is affected by stellar flares.\\
Before performing the flare analysis, all planetary transits were removed from the TESS light curves using literature ephemerides. For each planet, we adopted the published transit epoch, orbital period, and transit duration, and propagated the linear ephemeris across the TESS observing window. Transit mid-times were computed iteratively starting from the reference epoch and stepping forward and backward in time using the orbital period to cover the full time span of the observations. For each predicted transit, we removed data points within a window centered on the transit mid-time and extending to half the transit duration on either side. To account for uncertainties in the ephemerides and possible timing drifts over time, we applied an additional buffer by enlarging the exclusion window by 10\% of the transit duration. This ensures that ingress and egress phases, as well as potential deviations from the predicted timing, are fully excluded from the analysis.
This step is essential to avoid spurious flare detections caused by imperfect transit modeling causing residuals at ingress and egress, which could otherwise artificially inflate the inferred flare rates. The flare statistics are thus derived exclusively from out-of-transit stellar variability.\\
The probability estimation relies on the FFD of each star, modeled as a two-segment power law. In this work we assume that the energetic slope is an intrinsic estimate of the flare frequency. Using these fits, we estimate the expected rate of flares above a given energy threshold, expressed as a fraction of the stellar bolometric luminosity, $E_\mathrm{cut} = f \cdot L_\mathrm{bol}$. Two thresholds are considered in this work: a conservative threshold corresponding to very energetic events ($f = 1.0$, 100\% of $L_{bol}$) and a lower threshold representative of less intense but still potentially disruptive flares ($f = 0.1$, 10\% of $L_{bol}$).\\
To link the flare statistics to the observing conditions of Ariel, we define a temporal window representative of a typical transit visit. This window is set equal to twice the planetary transit duration, effectively assuming that flares occurring up to 50\% before ingress or 50\% after egress may still affect the observation through baseline distortions or biased detrending. An implicit assumption of this approach is that individual flare durations are generally shorter than the adopted window, which is justified for the majority of optical flares observed in the sample. Under this assumption, the probability of overlap between a flare and the transit window can be treated using a Poisson formalism.\\
For each star and energy threshold, the flare occurrence rate inferred from the appropriate segment of the FFD is multiplied by the observing window to obtain the expected number of flares that can contaminate the Ariel observation,
\begin{equation}
    \lambda = \text{rate} \times t_\mathrm{window}
\end{equation}
The probability of observing at least one flare during a single transit visit is then given by
\begin{equation}
    P_{flare}=(1-e^{-\lambda})\times 100
\end{equation}
Only stars with well-constrained two-segment FFD fits and with a measured minimum flare energy lower than the adopted threshold are included in this calculation. This criterion is satisfied by 184 stars when adopting a threshold of 10\% $L_{\mathrm{bol}}$, and by 229 stars when adopting a threshold of 100\% $L_{\mathrm{bol}}$. This ensures that the probability estimates rely on an interpolation anchored to the observed flare distribution rather than on an extrapolation beyond the energy range covered by the data.\\
In order to assess the robustness of the flare probability estimates, we quantified the level of agreement between the flare occurrence rates inferred from the fitted FFDs and those directly measured from the observed cumulative flare distributions. For each star and for each observing sector, we computed the relative percentage difference between the slope-predicted flare frequency at the adopted energy threshold, $f_\mathrm{slope}$, and the observed frequency derived from the cumulative flare distribution, $f_\mathrm{obs}$, evaluated at the same energy cut. The percentage difference is defined as
\begin{equation}
\Delta\% = \frac{|f_\mathrm{slope} - f_\mathrm{obs}|}{f_\mathrm{obs}} \times 100 ,
\end{equation}
and provides a direct diagnostic of how well the slope reproduces the empirical flare statistics in the energy range relevant for the probability calculation.\\
The observed frequency $f_\mathrm{obs}$ is obtained by interpolating the cumulative flare distribution in logarithmic energy space around the selected energy threshold, ensuring that the comparison is not biased by the discrete sampling of flare energies. The frequency $f_\mathrm{slope}$ is computed from the slope of the fitted FFD (below or above the break energy), consistently with the procedure adopted for the flare probability estimation. This comparison is performed independently for each sector, allowing us to capture possible sector-to-sector variations due to limited statistics, instrumental effects, or intrinsic stellar variability.\\
Large values of $\Delta\%$ indicate cases in which the FFD fit significantly overestimates or underestimates the observed flare rate at the relevant energy scale. In the context of transit contamination probabilities, an overestimation of the flare rate directly translates into an overestimated probability of flare occurrence during a transit visit. To limit the impact of such biases, we adopt a conservative quality criterion and exclude individual sectors for which $\Delta\% \geq 30\%$ from the probability analysis. This threshold represents a compromise between retaining a statistically meaningful sample and avoiding sectors where the inferred probabilities would be dominated by poorly constrained or unreliable extrapolations of the FFD.\\
Applying this criterion results in the exclusion of 103 sectors out of a total of 831 analyzed sectors ($\approx$12\%) for the more stringent threshold of 1.0 $L_\mathrm{bol}$. For the lower threshold of 0.1 $L_\mathrm{bol}$, fewer sectors are excluded: 53 out of 744 analyzed sectors ($\approx$7\%). 
These distributions of $\Delta\%$ for the two thresholds are shown in Fig. \ref{fig:hist_delta_percent_sector_1.0} (threshold = 1) and Fig. \ref{fig:hist_delta_percent_sector_0.1} (threshold = 0.1), highlighting that the fraction of excluded sectors is limited, indicating that the quality cut removes only the most discrepant sectors while preserving the majority of the data.\\
The effect of this selection depends on the number of available sectors per star.\\
Globally 0 stars are completely discarded from the final sample for either threshold. However, the number of available sectors is reduced for 11 stars in the 0.1 $L_\mathrm{bol}$ case and for 18 stars in the 1 $L_\mathrm{bol}$ case. The rejected sectors are spread along different stars, with no obvious clustering of bad sectors on few objects. Table \ref{tab:excluded_side_by_side} summarizes the number of excluded sectors relative to the total number of available sectors for stars with more than 1 excluded sector.\\
\begin{table}
 \caption{Stars with more than one excluded sector ($\Delta$\% $\geq$ 30\%) for the two thresholds. Left: threshold = 1 $L_\mathrm{bol}$. Right: threshold = 0.1 $L_\mathrm{bol}$.}
 \label{tab:excluded_side_by_side}
\begin{tabular}{lclc}
  \hline
  Star (1 $L_\mathrm{bol}$) & $Sect_{\rm excl}/Sect_{\rm tot}$ & Star (0.1 $L_\mathrm{bol}$) & $Sect_{\rm excl}/Sect_{\rm tot}$ \\
  \hline
  WASP-62 & 9/25 & HD23472 & 3/10 \\
  TOI-1670 & 6/33 & TOI-1670 & 3/33 \\
  TOI-201 & 5/27 & TOI-500 & 3/7 \\
  HD28109 & 4/34 & TOI-700 & 3/23 \\
  DS Tuc A & 4/5 & HD110113 & 2/3 \\
  L98-59 & 3/20 & HD191939 & 2/21 \\
  TOI-1450A & 3/30 & HD28109 & 2/34 \\
  HD23472 & 2/10 & TOI-1246 & 2/32 \\
  K2-266 & 2/3 & TOI-1450A & 2/30 \\
  KELT-20 & 2/6 & TOI-481 & 2/16 \\
  GJ357 & 2/4 & WASP-62 & 2/25 \\
  HD191939 & 2/21 &  &  \\
  TOI-1759 & 2/8 &  &  \\
  HD108236 & 2/3 &  &  \\
  TOI-481 & 2/16 &  &  \\
  TOI-500 & 2/7 &  &  \\
  TOI-700 & 2/23 &  &  \\
  TOI-712 & 2/24 &  &  \\
  \hline
 \end{tabular}
\end{table}
This approach ensures that the final flare probability estimates are based on sectors where the fitted FFD provides a faithful representation of the observed flare activity, while minimizing the risk of artificially inflating the contamination probabilities for Ariel transit observations. By explicitly quantifying and controlling the slope–data discrepancies, we improve the reliability of the inferred probabilities and strengthen the statistical foundation of the results presented in Fig. \ref{fig:flare_probability}.\\
Fig. \ref{fig:flare_probability} shows the distribution of flare probabilities for the two energy thresholds. For the most energetic events ($E_{cut}=1 L_\mathrm{bol}$), the vast majority of stars exhibit probabilities between 0\% and 5\%, indicating that such extreme flares are unlikely to occur during a single transit observation. When $E_{cut}=0.1 L_\mathrm{bol}$, the distribution broadens and a small subset of stars reaches substantially higher probabilities, exceeding 40\% in some cases. These objects, including AU Mic and HD 28109, are among the most active stars in the sample and contribute significantly to the high-probability tail.\\
Overall, this analysis shows that while most Ariel targets are unlikely to experience strong flare contamination during a single transit observation, a small number of highly active stars represent a non-negligible risk. The subset of targets with a flare contamination probability exceeding 20\% for the 0.1 $L_{\mathrm{bol}}$ threshold is reported in Table \ref{tab:high_risk_10}. These targets may require special attention in scheduling, data reduction, or interpretation, as stellar flares could represent a dominant source of astrophysical noise during transit observations. For the more stringent threshold of 1 $L_{\mathrm{bol}}$, only AU Mic exceeds this level.\\
Finally, it is worth noting that flare contamination is not limited to total emitted energy value. The duration and amplitude of individual events play a crucial role in determining their impact on transit observations, as short, intense flares may affect light curves differently than longer, moderate events. Although correlations between flare energy, duration, and amplitude are expected, a detailed exploration of these relationships lies beyond the scope of the present work and will be addressed in future studies.\\
Table 1 listing all systems for which a flare contamination probability could be computed. The table reports the stellar bolometric luminosity, the high-energy FFD slope, the index GF.01, and the flare contamination probabilities for the two adopted energy thresholds (1.0 and 0.1 $L_{\mathrm{bol}}$). The table is publicly available at \url{https://zenodo.org/records/18773384}.
\begin{table}
\caption{High-risk targets with flare contamination probability > 20\% for a threshold of 0.1 $L_{\mathrm{bol}}$.}
\label{tab:high_risk_10}
\begin{tabular}{lc|lc}
\hline
Target & Probability [\%] & Target & Probability [\%] \\
\hline
HD28109 d   & 59.1 & TOI-257 b    & 26.8 \\
HD28109 c   & 57.4 & DS Tuc A b     & 26.4 \\
AU Mic b     & 57.0 & TOI-444 b    & 26.4 \\
HD63935 c   & 37.7 & HD106315 c   & 26.4 \\
HD1397 b    & 36.4 & TOI-1136 b   & 25.6 \\
TOI-4641 b  & 35.3 & K2-266 c     & 25.5 \\
HIP67522 b  & 32.9 & HD63935 b    & 25.4 \\
KELT-11 b   & 32.6 & WASP-99 b    & 25.4 \\
TOI-1670 d  & 32.3 & TOI-2443 b   & 24.8 \\
HD221416 b  & 32.0 & TOI-712 d    & 24.2 \\
V1298Tau b  & 30.7 & HIP29442 b   & 24.0 \\
HD332231 b  & 29.7 & HD202772A b  & 23.8 \\
HAT-P-2 b   & 28.9 & HD89345 b    & 23.6 \\
HD191939 d  & 26.9 & TOI-561 e    & 22.9 \\
TOI-199 b   & 22.8 & TOI-712 e    & 22.7 \\
KELT-2A b   & 22.1 & TOI-1136 d   & 21.6 \\
TOI-1136 e  & 21.2 & HD191939 c   & 21.1 \\
TOI-1246 b  & 21.1 & K2-266 b     & 21.0 \\
TOI-622 b   & 20.8 & TOI-561 d    & 20.8 \\
\hline
\end{tabular}
\end{table}

\section{Discussion}\label{sec:Discussion}
In this section, we extend the analysis by comparing the results obtained for the Ariel target sample with those derived from other stellar populations and by examining specific case studies that provide additional physical context and validation of our statistical framework. In Sect.~\ref{sec:comparisonM}, we compare the Ariel targets with the M-dwarf sample of \cite{galletta2025exploring}, highlighting differences in flare activity and statistical trends across stellar populations. In Sect.~\ref{sec:aumic}, we first focus on AU Mic and DS Tuc as extreme cases of stellar activity, examining in detail their flare properties and the occurrence of flares within planetary transit events. We then extend the analysis to a broader and more representative sample of targets, selected to reflect the typical contamination regime of our full dataset. For this larger sample, we perform a systematic comparison between predicted flare contamination probabilities and direct TESS observations of transits. Taken together, these two complementary approaches allow us to validate the statistical flare probability framework both in extreme conditions and in the typical regime, demonstrating its ability to reproduce, within statistical uncertainties, the flare contamination rates observed in real transit data.
\subsection{Comparison with M stars sample}\label{sec:comparisonM}
\begin{figure*}
    \centering
    \subfloat[]{
        \includegraphics[width=0.32\hsize]{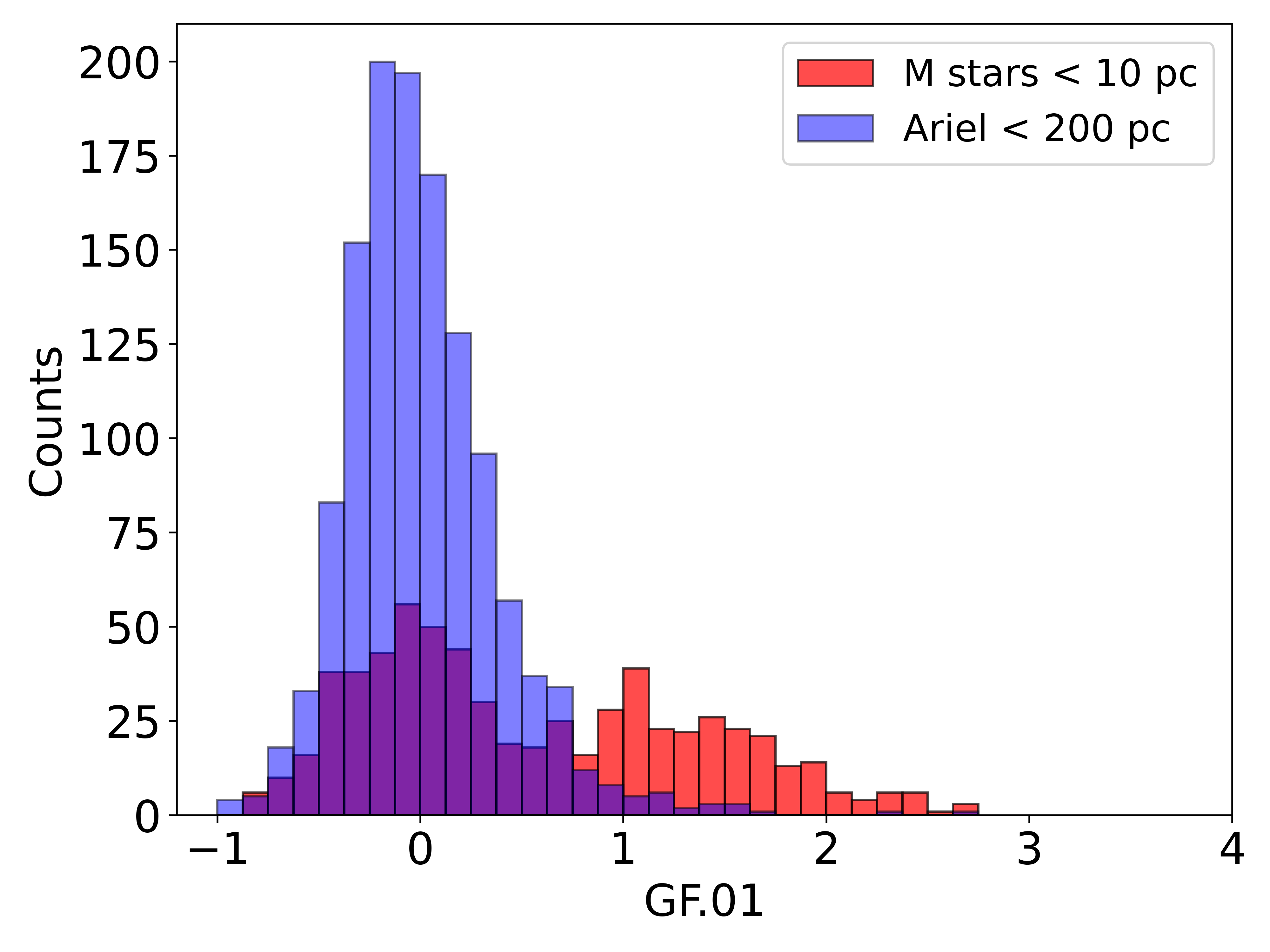}
        \label{fig:hist_gindex_ARIEL_vs_Mstars}
    }
    \subfloat[]{
        \includegraphics[width=0.32\hsize]{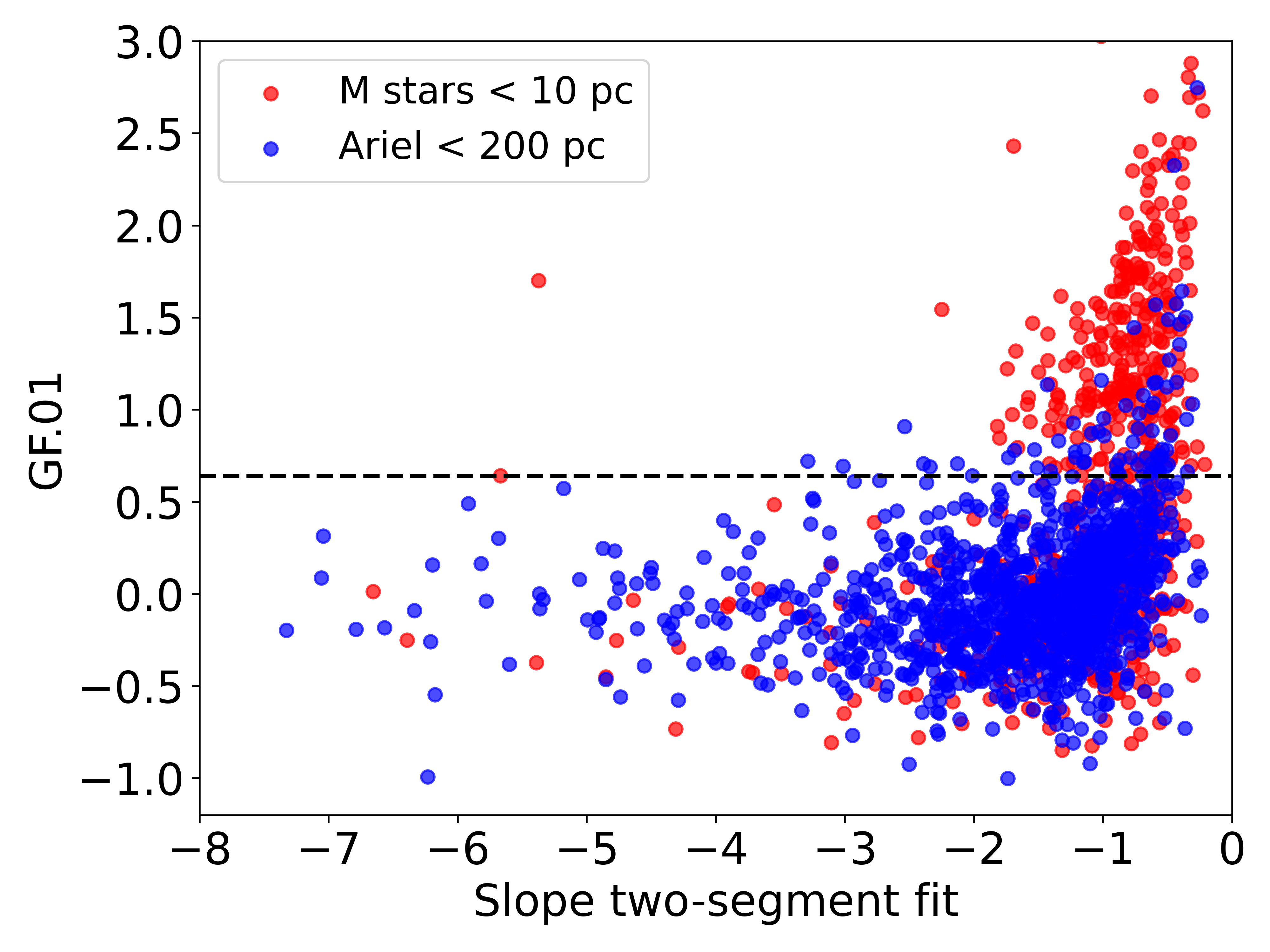}
        \label{fig:/slopevsgindex_ARIEL_vs_Mstars}
    }
    \subfloat[]{
        \includegraphics[width=0.32\hsize]{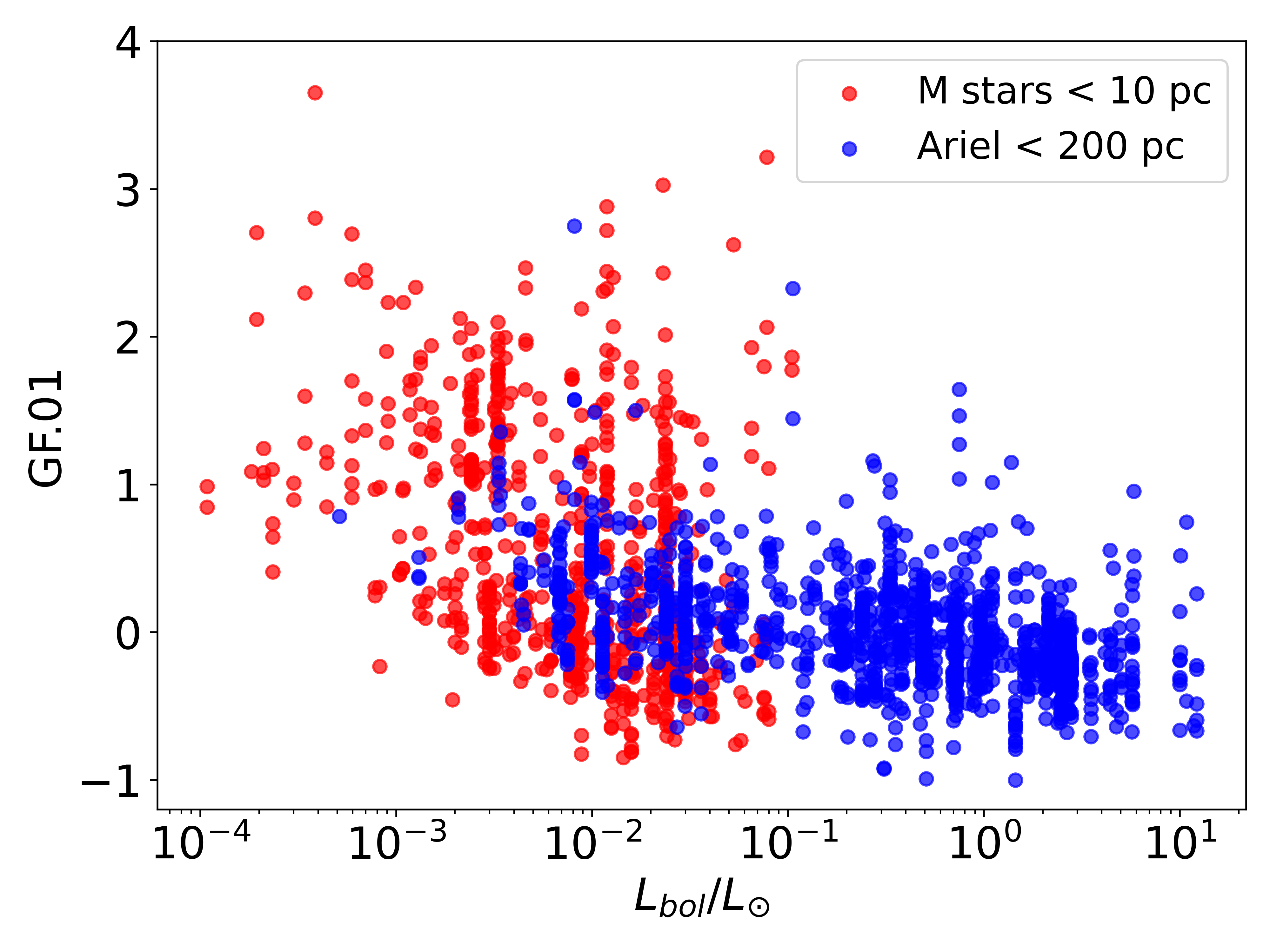}
        \label{fig:energyvsgindex_ARIEL_vs_Mstars}
    }
    \caption{(a) Distribution of the GF.01 index, defined as the flare energy corresponding to a frequency of 0.1 flares per day and normalized to the stellar bolometric luminosity. (b) GF.01 vs the high-energy slope of the two-segment flare frequency distribution. (c) GF.01 vs stellar bolometric luminosity ($L_{bol}$).
    In all panels, the blue points represent stars from the Ariel sample within 200 pc, while the red points correspond to M-dwarf stars from the sample of \protect\cite{galletta2025exploring} within 10 pc. This comparison highlights differences in flare activity properties between nearby low-mass stars and the broader Ariel target sample.}
    \label{fig:gindex_ARIEL_vs_M_stars}
\end{figure*}

To better understand how the flare properties of the Ariel targets compare with those of well-studied nearby active late-type stars, we performed a direct comparison between the Ariel sample and the M-dwarf sample analyzed by \cite{galletta2025exploring}. This comparison allows us to assess whether the trends observed in the local population of cool, magnetically active stars persist in the more diverse and distant set of stars selected for the Ariel mission.\\
Figure \ref{fig:gindex_ARIEL_vs_M_stars} presents the distributions and correlations of the GF.01 index for the two samples: the Ariel targets (blue, stars within 200 pc) and the M-dwarf sample (red, stars within 10 pc). The Ariel sample spans a wide range of spectral types and effective temperatures, whereas the nearby M-dwarf sample includes only cool stars below $\sim$ 4000 K. This distinction is clearly visible in Figure \ref{fig:hrdiagram}, where the dashed vertical black line marks the upper temperature boundary of M dwarfs. The Ariel stars extend beyond this regime. Furthermore, the Ariel target selection might introduce a bias against highly active stars, since strong activity complicates the detection and characterization of planets. This could partially explain the absence of very active stars in the Ariel dataset compared to the M-dwarf one.
The Fig. \ref{fig:hist_gindex_ARIEL_vs_Mstars} shows the distribution of the GF.01 index. The Ariel sample exhibits a single peak around zero with a modest positive tail. In contrast, the M-dwarf sample reveals a bimodal distribution, with one population centered near zero and another at higher GF.01 values, corresponding to a subset of highly active stars. This second peak, evident in the M dwarfs, is absent in the Ariel sample possibly due to selection effects or intrinsic differences in stellar types.
Fig. \ref{fig:/slopevsgindex_ARIEL_vs_Mstars} compares GF.01 with the high-energy slope from the two-segment power-law fit. Both samples follow a general trend in which higher flare activity, as measured by GF.01, corresponds to flatter slopes indicating that energetic flares become more frequent as overall activity increases. However the M-dwarf sample spans a wider range of GF.01 values and includes a greater number of highly active stars. The Ariel sample, instead, remains more tightly clustered around GF.01 $\approx$ 0.
Finally, Fig.\ref{fig:energyvsgindex_ARIEL_vs_Mstars} displays the relation between GF.01 and the stellar bolometric luminosity ($L_\mathrm{bol}$). 
As discussed for the Ariel sample (see Sect. \ref{sec:indexanalysis}), a negative trend is observed, with less luminous stars tending to exhibit higher relative flare activity.
For the M-dwarf sample, we also performed a linear fit, obtaining a slope of $m = -0.559 \pm 0.058$ and an intercept of $q = -0.608 \pm 0.128$. The fit yields a $\chi^2$ value of $0.689$, indicating an overall good agreement between the model and the observed distribution, indicating a statistically significant anti-correlation, steeper than the one observed in the Ariel dataset and in agreement with the results of \cite{galletta2025exploring}.\\
This behavior demonstrates that the relationship between stellar luminosity and relative flare strength extends beyond the nearby M dwarfs to the broader, more diverse Ariel target set. Overall, while the Ariel stars show a narrower range of activity levels, the consistency of these trends suggests that the fundamental connection between stellar energy output and magnetic activity is the same across different stellar populations.

\subsection{Extreme cases and method validation}\label{sec:aumic}
\begin{figure*}
    \centering
    \subfloat[]{
        \includegraphics[width=0.45\hsize]{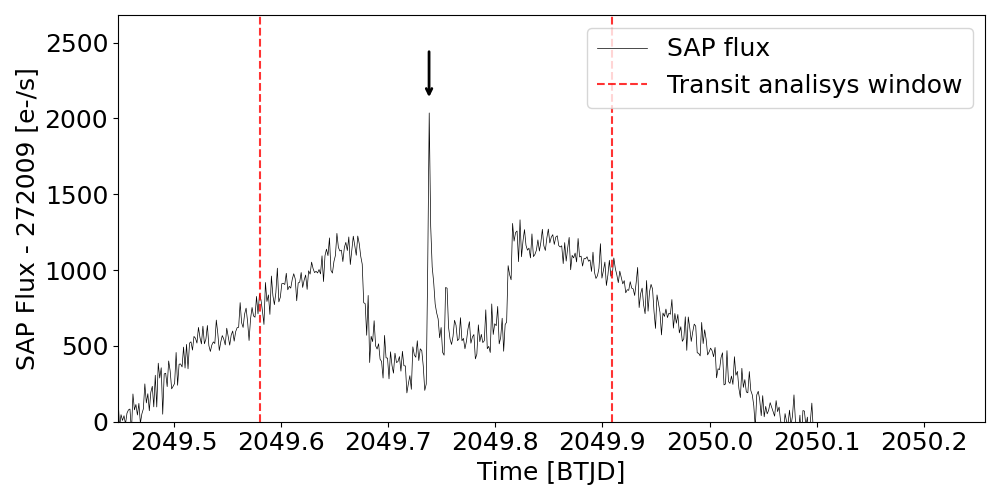}
        \label{fig:aumic_transit}
    }
    \subfloat[]{
        \includegraphics[width=0.45\hsize]{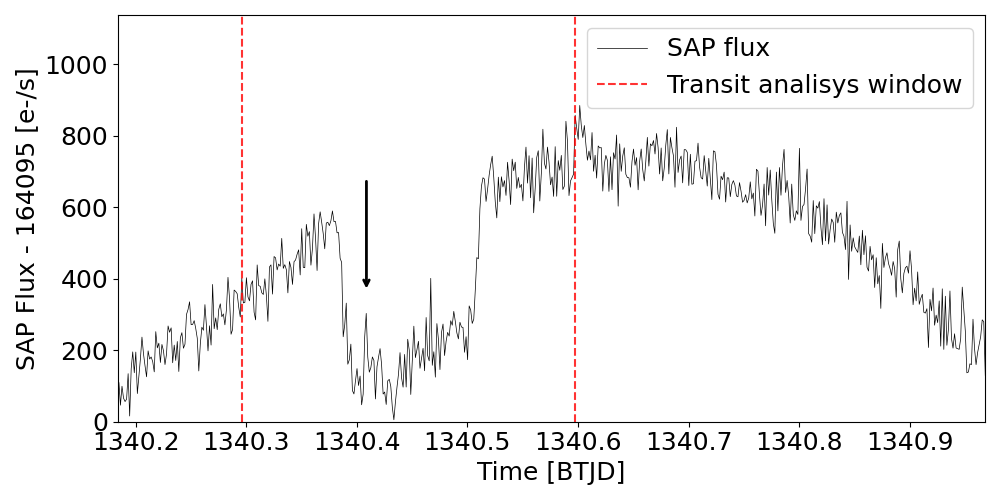}
        \label{fig:dstuc_transit}
    }

    \caption{Examples of stellar flare contamination during planetary transits observed by TESS. (a) Stellar flare occurring during a transit of AU Mic b, near the center of the transit. (b) Stellar flare occurring within the transit window of DS Tuc A b. In both panels, the red dashed vertical lines indicate the time interval adopted for the transit analysis, defined as the nominal transit duration extended by 50\% before ingress and 50\% after egress. The arrow marks the event identified as a stellar flare.}
    \label{fig:flare_inside_transit}
\end{figure*}

AU Mic emerges as an extreme case within our sample. For a flare energy threshold of 0.1 $L_{\mathrm{bol}}$, the estimated probability of observing at least one flare during a single transit event reaches approximately 57\%, making AU Mic the most flare-active target in our analysis. This exceptionally high probability reflects the intense and persistent magnetic activity of the star and places it at the upper end of the activity distribution among the analyzed targets.\\
The flare behavior of AU Mic has been extensively characterized in previous studies \citep[][]{ilin2022searching, feinstein2022microscopii, tristan20237}, notably by \cite{colombo2022short}, who presented its cumulative FFD and the statistical properties of flare amplitudes and decay times. Using the methodology developed by \cite{galletta2025exploring}, the present analysis achieves an increased sensitivity to low-energy flare events compared to previous approaches. This is primarily due to the flare-detection and fitting strategy, which is optimized to identify weak and short-lived events that would otherwise be difficult to detect. In particular, while previous studies \citep[e.g.][]{colombo2022short,martioli2021new} typically probe flare energies down to $\sim 10^{31}$ erg, our analysis reaches energies of $\sim 10^{30}$ erg. However, we emphasize that this low-energy regime corresponds to the biased-affected part of the flare frequency distribution, as also shown by our injection–recovery tests, and is therefore not used in the derivation of the main physical results of this work. This extension toward lower energies reflects an improved sensitivity of the detection pipeline, rather than a direct increase in the intrinsic flare occurrence rate.\\
However, while the enhanced detection of low-energy flares improves the completeness of the flare sample, these events do not play a dominant role in the context of the present work. In particular, they do not significantly contribute to the determination of the high-energy segment of the FFD, which is described by the second power-law slope and is the key quantity used to characterize the overall flare activity level relevant for planetary transit contamination. The second slope is primarily constrained by energetic flares above the break energy, and therefore remains largely unaffected by the improved sensitivity at low energies.\\
At high energies, AU Mic exhibits a remarkably high flare occurrence rate. From our analysis, the star produces about three flares per day with energies greater than or equal to $10^{32}$ erg, placing it among the most active known M dwarfs. This behavior is comparable to that observed in benchmark active stars such as EV Lac, which shows a similar flare rate at comparable energy thresholds \citep[][]{galletta2025exploring}. Such high-frequency, energetic flare activity indicates an efficient and sustained release of magnetic energy, consistent with AU Mic’s young age and the presence of a powerful magnetic dynamo.\\
The statistical properties of the flare population further reinforce this picture. In our analysis, the flares span a wide range of amplitudes and decay times, with events lasting up to approximately 1500 s. The distribution of amplitudes and timescales is broadly consistent with that observed in other highly active M dwarfs \citep[][]{galletta2025exploring}, although AU Mic tends to exhibit slightly shorter decay times than EV Lac. The presence of long-duration and energetic flares is particularly relevant for transit observations, as such events can overlap with planetary transits and significantly bias photometric and spectroscopic measurements.\\
Overall, AU Mic combines extremely high flare occurrence rates with energetic and, in some cases, long-duration events. Quantitatively, AU Mic exhibits an average slope of $-0.606 \pm 0.156$, together with a high flare rate indicator of $\mathrm{GF0.1} = 1.885 \pm 0.311$, placing it at the extreme end of the activity distribution within our sample.\\
A further validation of the extreme flare activity inferred from the statistical analysis is provided by a direct inspection of the available transit observations of AU Mic. The star was observed by TESS in two sectors, Sector 1 and Sector 27, during which a total of five planetary transits were observed. Remarkably, four out of these five transits are affected by at least one detectable stellar flare, corresponding to an empirical contamination rate of 80\%. This observational result is fully consistent with the high flare probability estimated from the FFD and provides an independent, event-level confirmation of the statistical predictions.\\
Examples of flare contamination during TESS transit observations are shown in Fig.~\ref{fig:flare_inside_transit}. Fig. \ref{fig:aumic_transit} illustrates a case for AU Mic b, where a flare occurs near the center of the transit. As already discussed, AU Mic represents an extreme and highly active system, in which strong flare contamination is relatively frequent. In this configuration, a flare overlapping the transit core produces a temporary increase in stellar flux that partially fills in the transit signal, leading to an underestimation of the transit depth and, consequently, of the inferred planet-to-star radius ratio.\\
However, the dataset also includes events that are more representative of the average flare contamination expected for the Ariel sample. An example is shown in Fig. ~\ref{fig:dstuc_transit} for DS Tuc A b, where a flare occurs within the transit analysis window but does not dominate the light curve as strongly as in AU Mic. This case illustrates a more typical scenario, in which flares contribute additional noise or mild distortions to the transit profile rather than completely altering its shape.\\
These examples highlight the diversity of flare–transit configurations present in the sample, ranging from extreme cases such as AU Mic to more moderate events. While strong contamination remains rare, lower-level flare activity within the transit window is not uncommon and should be accounted for in the analysis and interpretation of high-precision transit observations.\\
Beyond these illustrative cases, we also carried out a further check aimed at validating the flare contamination probabilities derived in this work against the available TESS observations. In particular, we selected a subset of systems characterized by a sufficiently large number of observed transits (at least 10) and intermediate predicted flare contamination probabilities in the range 20–30\%, about 10\% of the ARIEL sample. For these targets, we directly inspected all available transit windows in the TESS light curves, searching for the presence of flare-like events occurring during the transit intervals as defined in our analysis.\\
This procedure allowed us to obtain an empirical estimate of the fraction of transits affected by flaring activity, which we then compared with the probabilities predicted by our model. The results of this comparison are summarized in Tab. \ref{tab:probability_visual_flare}, where we report both the expected number of flare-contaminated transits predicted by the model and the number of transits in which a flare was visually identified in the TESS light curves. Overall, we find that the predicted and observed values are broadly consistent within statistical uncertainties. Importantly, when considering the full sample, the total number of observed flare-contaminated transits is the same to the total number expected from the model, with both yielding 84 events.
This agreement provides an important consistency check for our methodology, supporting the reliability of the flare contamination estimate beyond the extreme cases discussed previously, such as AU Mic. In particular, it indicates that the statistical approach used to infer flare occurrence during transit windows is able to reproduce, within the expected statistical scatter, the overall flare contamination rate observed in real TESS transit data, even if deviations are naturally expected for individual planetary systems.\\
This effect is particularly critical because it directly biases one of the fundamental planetary parameters. In addition, stellar flares are intrinsically chromatic, with a spectral energy distribution that is significantly bluer than the quiescent stellar photosphere. As a result, the apparent transit depth becomes wavelength-dependent, potentially mimicking or masking atmospheric signatures in transmission spectroscopy. In multi-band or spectroscopic observations, such chromatic contamination can introduce spurious slopes or features in the transmission spectrum, which may be incorrectly interpreted as evidence for atmospheric scattering, molecular absorption, or clouds.
\begin{table}
\centering
\caption{Comparison between predicted and observed flare contamination in planetary transit observations. The table lists systems with at least 10 observed transits and predicted flare contamination probabilities in the 20–30\% range. For each system, we report the number of analyzed transits, the expected and observed numbers of flare-contaminated transits, the model-predicted contamination probability per transit, and the observed contamination fraction derived from visual inspection of the TESS light curves.}
\label{tab:probability_visual_flare}
\resizebox{\columnwidth}{!}{%
\begin{tabular}{cccccc}
\hline
Planet & \# transit & \# $f_{expected}$ & \# $f_{observed}$ & $P_{model}$ (\%) & $P_{observed}$ (\%) \\ 
\hline
TOI-199 b   & 10 & $2 $   & $4 \pm 2.0$   & 23 & 40 \\
TOI-712 e   & 11 & $3$   & $2 \pm 1.4$   & 23 & 18 \\
HD63935 b   & 11 & $3$   & $1 \pm 1.0$   & 25 & 9  \\
TOI-1136 b  & 14 & $4$   & $3 \pm 1.7$   & 26 & 21 \\
TOI-712 d   & 15 & $4$   & $3 \pm 1.7$   & 24 & 20 \\
HD191939 d  & 18 & $5$   & $4 \pm 2.0$   & 27 & 22 \\
Kelt-11 b   & 22 & $7$   & $9 \pm 3.0$   & 33 & 41 \\
HD191939 c  & 24 & $5$   & $6 \pm 2.4$   & 21 & 25 \\
Wasp-99 b   & 25 & $6$   & $6 \pm 2.4$   & 25 & 24 \\
HAT-P-2 b   & 25 & $7$   & $7 \pm 2.6$   & 29 & 28 \\
TOI-622 b   & 31 & $7$   & $10 \pm 3.2$  & 21 & 32 \\
Kelt-2A b   & 35 & $8$   & $12 \pm 3.5$  & 22 & 34 \\
TOI-1246 b  & 50 & $11$  & $5 \pm 2.2$   & 21 & 10 \\
HD202772A b & 52 & $12$  & $12 \pm 3.5$  & 24 & 23 \\
\hline
Total & & $84$ & $84 \pm 9.2$ & & \\
\hline
\end{tabular}%
}
\end{table}
\section{Conclusions}\label{sec:conclusions}
In this work, we analyzed a sample of 290 Ariel target stars using TESS light curves to characterize stellar flare activity and its potential impact on transit observations. Building upon the methodology developed by \cite{colombo2022short} and updated by \cite{galletta2025exploring}, we identified individual flares, modeled their cumulative energy distributions with two-segment power laws, and derived high-energy slopes as indicators of stellar activity levels. 
Overall, the analysis led to the detection of 15857 flare events across 1638 analyzed TESS sectors, with the number of detected flares per sector ranging from 2 to 86 events.
The distribution of high-energy slopes across the sample shows a clear peak around –1.8, with a mean of –1.87 and a standard deviation of 1.55. Slopes $\leq$ –1 indicate that the flare energy distribution is dominated by low-energy events, implying that the overall flare activity is governed by a large number of frequent, low-energy flares rather than rare, high-energy ones. This behavior is typically associated with stars exhibiting relatively high quiescent brightness levels, where the magnetic energy release occurs more efficiently through numerous small-scale reconnection events, as explained in \cite{galletta2025exploring}.\\
By combining slope and energy break information with transit durations, we computed the probability of observing at least one flare, above an energy threshold, during a single transit for each planet. This analysis provides a robust and quantitative framework for assessing the likelihood of flare contamination, which is critical for planning Ariel observations and ensuring reliable atmospheric characterization.\\
Near 3\% of the sample exhibits enhanced flare activity (GF.01 > 1). AU Mic and HD 28109, shows a high probability of flare contamination during transit observations. In the case of AU Mic, this is driven by its intrinsically high flare rate. In contrast, for HD 28109 the elevated probability is primarily due to the long duration of its transits (exceeding 10 hours), which increases the likelihood of overlapping flare events despite its lower intrinsic activity level. A fraction of about 10\% of the sample has a contamination probability in 20\%-30\% range. 
A practical implication of this analysis concerns the observing strategy required to mitigate flare contamination. 
If $p$ denotes the probability that a transit is affected by at least one flare, the probability of obtaining a flare-free (``clean'') transit is given by $(1 - p)$. The number of observed transits required to secure a desired number $N_{\mathrm{clean}}$ of uncontaminated events therefore scales as:
\begin{equation}
N_{\mathrm{obs}} \simeq \frac{N_{\mathrm{clean}}}{1 - p}.
\end{equation}
This relation implies that the observational cost increases rapidly as the flare contamination probability increases. For example, if $p = 0.25$, about 25\% of transits are affected by flares, and the required number of observations increases by a factor of $\sim 1.33$. For $p = 0.5$, the required observing time doubles, while for $p = 0.75$ it increases by a factor of $\sim 4$, making such targets significantly more expensive in terms of observing time.\\
Therefore, the flare contamination probability derived in this work provides a quantitative criterion for target prioritization. Systems with high $p$ may require either a significant increase in allocated observing time or dedicated analysis strategies to model and correct for flare contamination.\\
Overall, our study demonstrates that detailed flare characterization enables reliable estimation of flare occurrence within transits, allowing for improved target selection, observation planning, and interpretation of exoplanet atmospheric signals, particularly for magnetically active low-mass stars.

\section*{Acknowledgements}

G. M. acknowledges support from the European Union - Next Generation EU through the grant n. 2022J7ZFRA - Exo-planetary Cloudy Atmospheres and Stellar High energy (Exo-CASH) funded by MUR - PRIN 2022 and the ASI-INAF agreement 2021-5-HH.2-2024.\\
This research made use of Lightkurve, a Python package for Kepler and TESS data analysis \citep{2018ascl.soft12013L}.\\
This publication makes use of VOSA, developed under the Spanish Virtual Observatory (\url{https://svo.cab.inta-csic.es}) project funded by MCIN/AEI/10.13039/501100011033/ through grant PID2020-112949GB-I00. VOSA has been partially updated by using funding from the European Union's Horizon 2020 Research and Innovation Programme, under Grant Agreement nº 776403 (EXOPLANETS-A).

\section*{CONFLICT OF INTEREST}
The authors declare no conflict of interest
\section*{Data Availability}
The data underlying this study are based on publicly available light curves from TESS, which can be accessed through the MAST archive (\url{https://mast.stsci.edu}).\\
The bolometric luminosities, two-segment FFD fit parameters, GF.01 activity indices, and the derived flare contamination probabilities for all Ariel targets analysed in this work are available in machine-readable format at \url{https://zenodo.org/records/18773384}.\\
All additional data products generated during this study, including intermediate flare lists and sector-level measurements, are available from the corresponding author upon reasonable request.



\bibliographystyle{rasti}
\bibliography{example} 

@article{radick1990stellar,
  title={Stellar activity and brightness variations: A glimpse at the Sun's history},
  author={Radick, Richard R and Lockwood, GW and Baliunas, Sallie L},
  journal={Science},
  volume={247},
  number={4938},
  pages={39--44},
  year={1990},
  publisher={American Association for the Advancement of Science}
}

@article{donati2008magnetic,
  title={Magnetic cycles of the planet-hosting star $\tau$ Bootis},
  author={Donati, J-F and Moutou, C and Far{\`e}s, R and Bohlender, D and Catala, C and Deleuil, M and Shkolnik, E and Cameron, A Collier and Jardine, Moira Mary and Walker, GAH},
  journal={MNRAS},
  volume={385},
  number={3},
  pages={1179--1185},
  year={2008},
  publisher={Blackwell Publishing Ltd Oxford, UK}
}

@article{jeffers2023stellar,
  title={Stellar activity cycles},
  author={Jeffers, Sandra V and Kiefer, Ren{\'e} and Metcalfe, Travis S},
  journal={Space Science Reviews},
  volume={219},
  number={7},
  pages={54},
  year={2023},
  publisher={Springer}
}

@article{lanza2009stellar,
  title={Stellar coronal magnetic fields and star-planet interaction},
  author={Lanza, AF},
  journal={A\&A},
  volume={505},
  number={1},
  pages={339--350},
  year={2009},
  publisher={EDP Sciences}
}

@article{cauley2019magnetic,
  title={Magnetic field strengths of hot Jupiters from signals of star--planet interactions},
  author={Cauley, P Wilson and Shkolnik, Evgenya L and Llama, Joe and Lanza, Antonino F},
  journal={Nature Astronomy},
  volume={3},
  number={12},
  pages={1128--1134},
  year={2019},
  publisher={Nature Publishing Group UK London}
}

@article{carmichael196454,
  title={A PROCESS FOR FLARES},
  author={Carmichael, HUGH},
  journal={NASA SP.},
  number={50},
  pages={451},
  year={1964},
  publisher={Scientific and Technical Information Office, National Aeronautics and Space~…}
}

@article{buccino2007uv,
  title={UV habitable zones around M stars},
  author={Buccino, Andrea P and Lemarchand, Guillermo A and Mauas, Pablo JD},
  journal={\icarus},
  volume={192},
  number={2},
  pages={582--587},
  year={2007},
  publisher={Elsevier}
}

@article{segura2010effect,
  title={The effect of a strong stellar flare on the atmospheric chemistry of an Earth-like planet orbiting an M dwarf},
  author={Segura, Ant{\'\i}gona and Walkowicz, Lucianne M and Meadows, Victoria and Kasting, James and Hawley, Suzanne},
  journal={Astrobiology},
  volume={10},
  number={7},
  pages={751--771},
  year={2010},
  publisher={Mary Ann Liebert, Inc. 140 Huguenot Street, 3rd Floor New Rochelle, NY 10801 USA}
}

@article{johnstone2016influences,
  title={The influences of stellar activity on planetary atmospheres},
  author={Johnstone, Colin P},
  journal={IAU Symposium},
  volume={12},
  number={S328},
  pages={168--179},
  year={2016},
  publisher={Cambridge University Press}
}

@article{oshagh2013effect,
  title={Effect of stellar spots on high-precision transit light-curve},
  author={Oshagh, Mahmoud and Santos, Nuno C and Boisse, Isabelle and Boue, Gwenael and Montalto, Marco and Dumusque, Xavier and Haghighipour, Nader},
  journal={Astronomy \& Astrophysics},
  volume={556},
  pages={A19},
  year={2013},
  publisher={EDP Sciences}
}

@inproceedings{tinetti2016science,
  title={The science of ARIEL (atmospheric remote-sensing infrared exoplanet large-survey)},
  author={Tinetti, Giovanna and Drossart, Pierre and Eccleston, Paul and Hartogh, Paul and Heske, A and Leconte, J and Micela, G and Ollivier, M and Pilbratt, G and Puig, L and others},
  booktitle={Space Telescopes and Instrumentation 2016: Optical, Infrared, and Millimeter Wave},
  volume={9904},
  pages={658--667},
  year={2016},
  organization={SPIE}
}

@article{tinetti2018chemical,
  title={A chemical survey of exoplanets with ARIEL},
  author={Tinetti, Giovanna and Drossart, Pierre and Eccleston, Paul and Hartogh, Paul and Heske, Astrid and Leconte, J{\'e}r{\'e}my and Micela, Giusi and Ollivier, Marc and Pilbratt, G{\"o}ran and Puig, Ludovic and others},
  journal={Experimental astronomy},
  volume={46},
  number={1},
  pages={135--209},
  year={2018},
  publisher={Springer}
}

@INPROCEEDINGS{2022EPSC...16.1114T,
       author = {{Tinetti}, Giovanna and {Eccleston}, Paul and {Lueftinger}, Theresa and {Salvignol}, Jean-Christophe and {Fahmy}, Salma and {Alves de Oliveira}, Caterina},
        title = "{Ariel: Enabling planetary science across light-years}",
    booktitle = {European Planetary Science Congress},
         year = 2022,
        month = sep,
          eid = {EPSC2022-1114},
        pages = {EPSC2022-1114},
          doi = {10.5194/epsc2022-1114},
archivePrefix = {arXiv},
       eprint = {2104.04824},
 primaryClass = {astro-ph.IM},
       adsurl = {https://ui.adsabs.harvard.edu/abs/2022EPSC...16.1114T}
}

@article{galletta2025exploring,
  title={Exploring short-term stellar activity in M dwarfs: A volume-limited perspective},
  author={Galletta, G and Colombo, S and Prisinzano, L and Micela, G},
  journal={Astronomy \& Astrophysics},
  volume={698},
  pages={A180},
  year={2025},
  publisher={EDP Sciences}
}

@inproceedings{ricker2016transiting,
  title={The transiting exoplanet survey satellite},
  author={Ricker, George R and Vanderspek, R and Winn, J and Seager, S and Berta-Thompson, Z and Levine, A and Villasenor, J and Latham, D and Charbonneau, D and Holman, M and others},
  booktitle={Space Telescopes and Instrumentation 2016: Optical, Infrared, and Millimeter Wave},
  volume={9904},
  pages={767--784},
  year={2016},
  organization={SPIE}
}

@article{edwards2022ariel,
  title={The Ariel target list: The impact of TESS and the potential for characterizing multiple planets within a system},
  author={Edwards, Billy and Tinetti, Giovanna},
  journal={The Astronomical Journal},
  volume={164},
  number={1},
  pages={15},
  year={2022},
  publisher={IOP Publishing}
}

@article{mugnai2020arielrad,
  title={ArielRad: the Ariel radiometric model},
  author={Mugnai, Lorenzo V and Pascale, Enzo and Edwards, Billy and Papageorgiou, Andreas and Sarkar, Subhajit},
  journal={Experimental Astronomy},
  volume={50},
  number={2},
  pages={303--328},
  year={2020},
  publisher={Springer}
}

@article{edwards2019updated,
  title={An updated study of potential targets for Ariel},
  author={Edwards, Billy and Mugnai, Lorenzo and Tinetti, Giovanna and Pascale, Enzo and Sarkar, Subhajit},
  journal={The Astronomical Journal},
  volume={157},
  number={6},
  pages={242},
  year={2019},
  publisher={American Astronomical Society}
}

@MISC{2018ascl.soft12013L,
   author = {{Lightkurve Collaboration} and {Cardoso}, J.~V.~d.~M. and
             {Hedges}, C. and {Gully-Santiago}, M. and {Saunders}, N. and
             {Cody}, A.~M. and {Barclay}, T. and {Hall}, O. and
             {Sagear}, S. and {Turtelboom}, E. and {Zhang}, J. and
             {Tzanidakis}, A. and {Mighell}, K. and {Coughlin}, J. and
             {Bell}, K. and {Berta-Thompson}, Z. and {Williams}, P. and
             {Dotson}, J. and {Barentsen}, G.},
    title = "{Lightkurve: Kepler and TESS time series analysis in Python}",
howpublished = {Astrophysics Source Code Library},
     year = 2018,
    month = dec,
archivePrefix = "ascl",
   eprint = {1812.013},
   adsurl = {http://adsabs.harvard.edu/abs/2018ascl.soft12013L},
}

@inproceedings{rodrigo2020vosa,
  title={VOSA 7.0-A VO Spectral Energy Distribution Analyzer. New features},
  author={Rodrigo, C and Bayo Ar{\'a}n, A and Solano, E and Cort{\'e}s-Contreras, M},
  booktitle={XIV. 0 Scientific Meeting (Virtual) of the Spanish Astronomical Society},
  pages={181},
  year={2020}
}

@article{kowalski2024stellar,
  title={Stellar flares},
  author={Kowalski, Adam F},
  journal={Living Reviews in Solar Physics},
  volume={21},
  number={1},
  pages={1},
  year={2024},
  publisher={Springer}
}

@article{colombo2022short,
  title={Short-term variability of DS Tucanae A observed with TESS},
  author={Colombo, Salvatore and Petralia, Antonio and Micela, Giuseppina},
  journal={Astronomy \& Astrophysics},
  volume={661},
  pages={A148},
  year={2022},
  publisher={EDP Sciences}
}

@article{ballerini2012multiwavelength,
  title={Multiwavelength flux variations induced by stellar magnetic activity: effects on planetary transits},
  author={Ballerini, P and Micela, G and Lanza, AF and Pagano, I},
  journal={Astronomy \& Astrophysics},
  volume={539},
  pages={A140},
  year={2012},
  publisher={EDP Sciences}
}

@article{ilin2022searching,
  title={Searching for flaring star--planet interactions in AU Mic TESS observations},
  author={Ilin, Ekaterina and Poppenh{\"a}ger, Katja},
  journal={Monthly Notices of the Royal Astronomical Society},
  volume={513},
  number={3},
  pages={4579--4586},
  year={2022},
  publisher={Oxford University Press}
}

@article{feinstein2022microscopii,
  title={AU Microscopii in the Far-UV: Observations in Quiescence, during Flares, and Implications for AU Mic b and c},
  author={Feinstein, Adina D and France, Kevin and Youngblood, Allison and Duvvuri, Girish M and Teal, DJ and Cauley, P Wilson and Seligman, Darryl Z and Gaidos, Eric and Kempton, Eliza M-R and Bean, Jacob L and others},
  journal={The Astronomical Journal},
  volume={164},
  number={3},
  pages={110},
  year={2022},
  publisher={The American Astronomical Society}
}

@article{tristan20237,
  title={A 7 day multiwavelength flare campaign on AU Mic. I. High-time-resolution light curves and the thermal empirical Neupert effect},
  author={Tristan, Isaiah I and Notsu, Yuta and Kowalski, Adam F and Brown, Alexander and Wisniewski, John P and Osten, Rachel A and Vrijmoet, Eliot H and White, Graeme L and Carter, Brad D and Grady, Carol A and others},
  journal={The Astrophysical Journal},
  volume={951},
  number={1},
  pages={33},
  year={2023},
  publisher={The American Astronomical Society}
}

@article{rackham2018transit,
  title={The transit light source effect: false spectral features and incorrect densities for M-dwarf transiting planets},
  author={Rackham, Benjamin V and Apai, D{\'a}niel and Giampapa, Mark S},
  journal={The Astrophysical Journal},
  volume={853},
  number={2},
  pages={122},
  year={2018},
  publisher={The American Astronomical Society}
}

@article{rackham2017access,
  title={Access i. an optical transmission spectrum of gj 1214b reveals a heterogeneous stellar photosphere},
  author={Rackham, Benjamin and Espinoza, N{\'e}stor and Apai, D{\'a}niel and L{\'o}pez-Morales, Mercedes and Jord{\'a}n, Andr{\'e}s and Osip, David J and Lewis, Nikole K and Rodler, Florian and Fraine, Jonathan D and Morley, Caroline V and others},
  journal={The Astrophysical Journal},
  volume={834},
  number={2},
  pages={151},
  year={2017},
  publisher={The American Astronomical Society}
}

@article{howard2023characterizing,
  title={Characterizing the near-infrared spectra of flares from TRAPPIST-1 during JWST transit spectroscopy observations},
  author={Howard, Ward S and Kowalski, Adam F and Flagg, Laura and MacGregor, Meredith A and Lim, Olivia and Radica, Michael and Piaulet, Caroline and Roy, Pierre-Alexis and Lafreni{\`e}re, David and Benneke, Bj{\"o}rn and others},
  journal={The Astrophysical Journal},
  volume={959},
  number={1},
  pages={64},
  year={2023},
  publisher={The American Astronomical Society}
}

@article{shibayama2013superflares,
  title={Superflares on solar-type stars observed with Kepler. I. Statistical properties of superflares},
  author={Shibayama, Takuya and Maehara, Hiroyuki and Notsu, Shota and Notsu, Yuta and Nagao, Takashi and Honda, Satoshi and Ishii, Takako T and Nogami, Daisaku and Shibata, Kazunari},
  journal={The Astrophysical Journal Supplement Series},
  volume={209},
  number={1},
  pages={5},
  year={2013},
  publisher={The American Astronomical Society}
}

@article{davenport2014kepler,
  title={Kepler flares. II. The temporal morphology of white-light flares on GJ 1243},
  author={Davenport, James RA and Hawley, Suzanne L and Hebb, Leslie and Wisniewski, John P and Kowalski, Adam F and Johnson, Emily C and Malatesta, Michael and Peraza, Jesus and Keil, Marcus and Silverberg, Steven M and others},
  journal={The Astrophysical Journal},
  volume={797},
  number={2},
  pages={122},
  year={2014},
  publisher={The American Astronomical Society}
}

@article{gunther2020stellar,
  title={Stellar flares from the first TESS data release: exploring a new sample of M dwarfs},
  author={G{\"u}nther, Maximilian N and Zhan, Zhuchang and Seager, Sara and Rimmer, Paul B and Ranjan, Sukrit and Stassun, Keivan G and Oelkers, Ryan J and Daylan, Tansu and Newton, Elisabeth and Kristiansen, Martti H and others},
  journal={The Astronomical Journal},
  volume={159},
  number={2},
  pages={60},
  year={2020},
  publisher={The American Astronomical Society}
}

@article{howard2019evryflare,
  title={EvryFlare. I. Long-term evryscope monitoring of flares from the cool stars across half the southern sky},
  author={Howard, Ward S and Corbett, Hank and Law, Nicholas M and Ratzloff, Jeffrey K and Glazier, Amy and Fors, Octavi and Ser, Daniel del and Haislip, Joshua},
  journal={The Astrophysical Journal},
  volume={881},
  number={1},
  pages={9},
  year={2019},
  publisher={The American Astronomical Society}
}

@article{howard2022flaring,
  title={The flaring TESS Objects of Interest: flare rates for all two-minute cadence TESS planet candidates},
  author={Howard, Ward S},
  journal={Monthly Notices of the Royal Astronomical Society: Letters},
  volume={512},
  number={1},
  pages={L60--L65},
  year={2022},
  publisher={Oxford University Press}
}

@article{ilin2021flares,
  title={Flares in open clusters with K2-II. Pleiades, Hyades, Praesepe, Ruprecht 147, and M 67},
  author={Ilin, Ekaterina and Schmidt, Sarah J and Poppenh{\"a}ger, Katja and Davenport, James RA and Kristiansen, Martti H and Omohundro, Mark},
  journal={Astronomy \& Astrophysics},
  volume={645},
  pages={A42},
  year={2021},
  publisher={EDP Sciences}
}

@article{martioli2021new,
  title={New constraints on the planetary system around the young active star AU Mic-Two transiting warm Neptunes near mean-motion resonance},
  author={Martioli, E and H{\'e}brard, G and Correia, ACM and Laskar, J and Des Etangs, A Lecavelier},
  journal={Astronomy \& Astrophysics},
  volume={649},
  pages={A177},
  year={2021},
  publisher={EDP Sciences}
}

@article{hawley2014kepler,
  title={Kepler flares. I. Active and inactive M dwarfs},
  author={Hawley, Suzanne L and Davenport, James RA and Kowalski, Adam F and Wisniewski, John P and Hebb, Leslie and Deitrick, Russell and Hilton, Eric J},
  journal={The Astrophysical Journal},
  volume={797},
  number={2},
  pages={121},
  year={2014},
  publisher={The American Astronomical Society}
}

@article{davenport2016kepler,
  title={The Kepler catalog of stellar flares},
  author={Davenport, James RA},
  journal={The Astrophysical Journal},
  volume={829},
  number={1},
  pages={23},
  year={2016},
  publisher={The American Astronomical Society}
}

@article{aschwanden2021self,
  title={Self-organized criticality in stellar flares},
  author={Aschwanden, Markus J and G{\"u}del, Manuel},
  journal={The Astrophysical Journal},
  volume={910},
  number={1},
  pages={41},
  year={2021},
  publisher={The American Astronomical Society}
}

@article{capistrant2026stellar,
  title={Stellar Flares in the TESS Light Curves of Planet-hosting M dwarfs},
  author={Capistrant, Benjamin K and Dittmann, Jason},
  journal={The Astrophysical Journal},
  volume={998},
  number={1},
  pages={173},
  year={2026},
  publisher={The American Astronomical Society}
}

@article{rajpurohit2025exploring,
  title={Exploring stellar activity in a sample of active M dwarfs},
  author={Rajpurohit, AS and Kumar, V and Srivastava, MK and Labadie, L and Rajpurohit, K and Fern{\'a}ndez-Trincado, JG},
  journal={Astronomy \& Astrophysics},
  volume={704},
  pages={A154},
  year={2025},
  publisher={EDP Sciences}
}

@article{foreman2017fast,
  title={Fast and scalable Gaussian process modeling with applications to astronomical time series},
  author={Foreman-Mackey, Daniel and Agol, Eric and Ambikasaran, Sivaram and Angus, Ruth},
  journal={The Astronomical Journal},
  volume={154},
  number={6},
  pages={220},
  year={2017},
  publisher={The American Astronomical Society}
}





\bsp	
\label{lastpage}
\end{document}